\documentclass[twocolumn,superscriptaddress,floatfix,prb,amsmath,aps,showpacs]{revtex4}
\usepackage{graphicx}
\usepackage{bm}
\usepackage{amssymb}
\usepackage[dvips]{color}
\begin{document}
%\floatsep=1cm
\bibliographystyle{apsrev}
\def\nn{\nonumber}
\def\dag{\dagger}
\def\u{\uparrow}
\def\d{\downarrow}
\def\j{\bm b}
\def\m{\bm m}
\def\l{\bm l}
\def\0{\bm 0}
\def\k{\bm k}
\title{Magnetostatic wave analog of integer quantum 
Hall state in patterned magnetic films}
\date{\today}
\author{Ryuichi Shindou} 
\affiliation{International Center for Quantum Materials, Peking 
University, No.5 Yiheyuan Road, Haidian District, Beijing, 100871, China}
\affiliation{Collaborative Innovation Center of Quantum
Matter, Beijing, China} 
\affiliation{Department of Physics, Tokyo Institute of Technology,
2-12-1 Ookayama, Meguro-ku, Tokyo, 152-8551, Japan} 
\author{Jun-ichiro Ohe}
\affiliation{Department of Physics, Toho University, 
2-2-1 Miyama, Funabashi, Chiba, Japan}
\begin{abstract}
A magnetostatic spin wave analog of integer quantum Hall (IQH) 
state is proposed in realistic patterned 
ferromagnetic thin films. Due to magnetic shape anisotropy,   
magnetic moments in a thin film lie within the plane, while   
all spin-wave excitations are fully gapped. Under 
an out-of-plane magnetic field, the film acquires a finite 
magnetization, where some of the gapped magnons 
become significantly softened near a saturation field. 
It is shown that,   
owing to a spin-orbit locking nature of the magnetic dipolar 
interaction, these soft spin-wave volume-mode bands 
become chiral volume-mode bands with 
finite topological Chern integers. A bulk-edge 
correspondence in IQH physics suggests  
that such volume-mode bands are accompanied 
by a chiral magnetostatic spin-wave 
edge mode. The existence of the edge 
mode is justified both by micromagnetic simulations and 
by band calculations based on a linearized 
Landau-Lifshitz equation. Employing 
intuitive physical arguments, we introduce proper 
tight-binding models for these soft volume-mode bands. 
Based on the tight-binding models, we further 
discuss possible applications to other systems 
such as magnetic ultrathin films with perpendicular magnetic 
anisotropy (PMA). 
\end{abstract}
\maketitle
%%%%%%%%%%%%

\section{introduction}
Spin-wave propagations in magnetic insulators realize 
spin transports with less dissipation,~\cite{YIG,KDG} fostering 
much prospect for realizations of future spintronic devices. 
For the purpose of device applications, spin-wave 
transport in two-dimensional systems such as 
thin films is expected to have many advantages. 
In ferromagnet thin film, moments lie within the 
plane to minimize the magnetostatic energy. 
A thin film with the 
in-plane magnetization has a 
surface spin-wave mode called Damon-Eshbach (DE) 
mode,~\cite{DE1} where spin wave propagates in a chiral 
direction transverse to the in-plane moment.   
The mode realizes a unidirectional spin transport 
in the two-dimensional (2-$d$) top surface of the film and 
the counter-propagating transport in the bottom 
surface. The mode enables a number of spin-wave  
spintronic devices.~\cite{Kostylev,Lee,Schneider,Sato} 

Recently, the present author proposes chiral spin-wave edge 
mode in a 2-$d$ {\it periodically-structured}  
dipolar magnetic thin film with {\it out-plane} ferromagnetic 
moment.~\cite{SO1,SO2} 
The mode has a resonance frequency within a band gap of 
volume modes, where the gap and multiple-band structure 
of volume-mode bands come from the 2-$d$ 
periodic structuring. 
The chiral direction is transverse to the out-of-plane 
ferromagnetic moment; the mode realizes a unidirectional 
spin-wave propagation along the one-dimensional 
boundary of the plane, instead of along the top (or bottom) 
surface. Such chiral edge modes could possibly connect 
various elements in 2-$d$ spin-current circuits in more flexible 
way than the DE surface mode. Moreover, the chiral 
direction (whether clockwise or counterclockwise) 
and number of the edge modes 
(can be more than one) are 
determined by a sum of the topological number (Chern integer)  
defined for the volume-mode bands below 
the gap.~\cite{TKKN,Hal,Hat,SO1,SO2} 
This enables us to control the 
direction and number of the edge modes in terms 
of a band gap manipulation, bringing up further 
prospect for spin-current circuits with richer structures.~\cite{SO1} 
To make such spin-wave circuits experimentally, 
it is much more important for theory 
to propose a number of structured thin 
films which have these topological modes.

In this paper, we introduce an efficient method  
of constructing the topological chiral edge 
modes in realistic dipolar magnetic thin films. 
We considered that magnetic clusters, either 
thin rings or circular disks, form a 2-$d$ periodic square 
lattice. To study their spin-wave excitations, we derive 
several tight-binding models, using intuitive physical 
arguments. Based on these models, we show that 
soft volume-mode bands near the saturation field acquire finite 
topological Chern integer, resulting in chiral spin wave 
edge modes within band gaps of volume mode bands. 
 
\begin{figure}[t]
   \includegraphics[width=70mm]{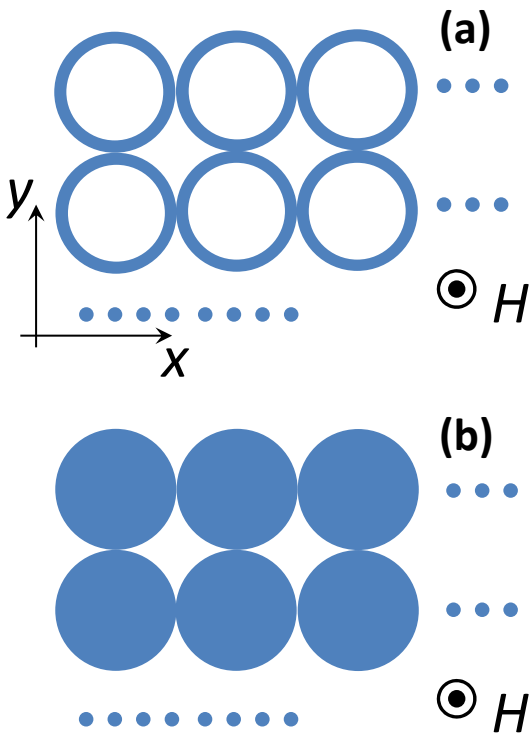}
\caption{ (Color online) Schematic top-view of two-dimensional 
patterned magnetic thin films (blue region represents 
magnetic media, while the other stands for the vacuum). 
An external magnetic field is applied perpendicular to the 
plane with out-of-plane moments. Either circular rings (a) or disks 
(b) form a square lattice. We assume that the film is 
sufficiently thin, so that 
there is no texture along the direction 
perpendicular to the plane.} 
\label{fig:cluster0}
\end{figure}

The organization of the paper is as follows. 
In sec.~II, we consider a ring model; 
circular magnetic thin rings forming a square lattice 
(Fig.~\ref{fig:cluster0}(a)). We first introduce 
an `atomic-orbital' like wavefunction for spin-wave excitations 
within each ring. Using these atomic orbitals, we  
construct tight-binding models for soft magnons. 
The models naturally lead to chiral volume-mode bands 
and edge modes (Fig.~\ref{fig:1.2-1} and Fig.~\ref{fig:1.5-2}).  
In sec.~III, we further extend the argument 
to a disk model, circular magnetic disks forming a 
square lattice (Fig.~\ref{fig:cluster0}(b)). The same type of 
chiral spin-wave edge modes are shown to appear 
in low-frequency regions  
near the saturation field (Fig.~\ref{fig:cir-band}). 
To justify the existence of the chiral edge modes 
by a standard method in the field, we also carried out 
in sec.~IV micromagnetic simulations in the 
proposed magnetic superlattices (Fig.~\ref{fig:mc3}).  In sec. V, 
we further discuss possible application of the present 
theory to other systems such as ferromagnetic ultrathin 
film systems with the perpendicular magnetic anisotropy.
Two appendices describe some details useful for understanding 
the main text. In the appendix A, we describe 
how wavelength-frequency dispersion relations for spin-wave 
volume-mode bands and edge mode bands (such as 
Fig.~\ref{fig:1.2-1}, Fig.~\ref{fig:1.5-2} and 
Fig.~\ref{fig:cir-band}) are calculated  from Landau-Lifshitz 
equations. In appendix B, we construct, in a more expliclit way, 
an effective tight-binding model for soft spin-wave 
excitations above the saturation field, which is helpful 
for understanding sec. II, and sec.IV in detail.        

All the results presented in this paper 
are essentially scalable, since the models do not have 
any short-range exchange interactions; the saturation 
field, $H_c$, and spin-wave resonance frequency 
are scaled only by the saturation magnetization 
(per volume) $M_s$ (appendix A). We took $M_s$ to be 
typically on the order of unit in Fig.~\ref{fig:ring-atlv}, 　
Fig.~\ref{fig:1.2-1}, Fig.~\ref{fig:1.5-2}, 
Fig.~\ref{fig:cir-atlv} and Fig.~\ref{fig:cir-band}, 
while it is on the order of GHz (see e.g. Sec.IV).

\section{ring model}
To begin with, consider spin-wave excitations in a 
magnetic circular ring. When a linear dimension of a  
cross section of the ring is 
comparable to short-ranged exchange length $l_{\rm ex}$ of a 
constituent magnetic material, 
the ring may be treated as a one-dimensional 
chain of $M$ spins, which are coupled with one another via 
long-range dipole-dipole interaction. $M$ is the number of 
the spins along the ring and is on the order of 
$2\pi r/l_{\rm ex}$ ($r$ is the radius of the ring). 
Without the field, the magnetostatic 
energy is minimized by a vortex spin 
configuration: spins are aligned along the tangential 
direction of the ring. Under the out-of-plane 
magnetic field $H$, the vortex spin configuration acquires an  
out-of-plane moment which becomes fully polarized above 
the saturation field, $H>H_c$. 

Suppose that the amplitude 
of each spin moment is fixed to be $M_s$. 
Excitations in each spin comprise two real-valued 
fields (transverse moments), so that the ring 
has $M$ numbers of complex-valued spin-wave modes, 
$\psi(\theta_j)$ with $\theta_j\equiv 2\pi j /M$ 
($j=1,\cdots,M$). Under a proper gauge choice  
(see appendix A), they have total angular 
momentum $q_J$ as their quantum number, 
\begin{align}
\psi_{q_J}(\theta_j+\theta_m)=e^{iq_Jm}\psi_{q_J}(\theta_j), \label{eq2}
\end{align} 
which comes from the circular rotational symmetry of a ring. 
Here $\theta_j\equiv 2\pi j /M$ and $q_J\equiv 2\pi n_J/M$ 
$(n_J=-M/2,-M/2+1,\cdots,M/2)$. The resonance frequency for 
these `atomic orbitals' is given as a function of the 
angular momentum, which forms a frequency 
band for larger $M$.  

\begin{figure}[t]
   \includegraphics[width=85mm]{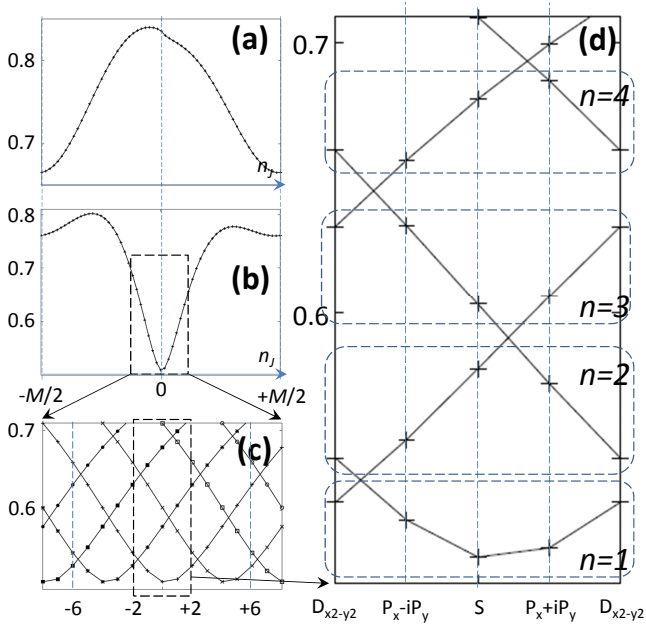}
\caption{ (Color online) Resonance frequency levels 
in a magnetic ring (a $M$-spins chain with $M=60$) 
as a function of the angular 
momentum $q_J \equiv \frac{2\pi n_J}{M}$ with 
$n_J=-M/2,-M/2+1,\cdots,M/2$. (a) $H=0.7H_c$ and 
(b) $H=0.9H_c$. (c) When demagnetization fields  
from the surrounding magnetic rings are included, 
the angular momentum $n_J$ is defined mod 
$4$. (d) Wave functions with $n_J \equiv$ $-1$,$0$,$+1$, $+2$ 
(mod $4$) are referred to as $P_-$,$S$,$P_+$ $D_{x^2-y^2}$-wave 
respectively, since they acquire $-i$, $+1$, $+i$ and $-1$ 
phase under the $\frac{\pi}{2}$ spatial rotation 
(eqs.~(\ref{s-w},\ref{p-w},\ref{d-w})).}
\label{fig:ring-atlv}
\end{figure}

At the zero field, all the spins in the circular vortex 
is along the angular momentum axis (along the tangential  
direction of the ring), so that the frequency band at $H=0$ becomes 
essentially same as the `backward'  volume modes  
in an in-plane magnetized thin film~\cite{DV2,Kalinikos} 
or cylindrically magnetized nanowire.~\cite{AM} Namely, the 
band has its resonance frequency minimum at 
$q_J=\pi$ and its frequency maximum at $q_J=0$. When 
increasing the out-of-plane field, the maximum and minimum 
are inverted at some `critical' field 
below the saturation field, 
$H=H_d\simeq 0.8H_c$ (Fig.~\ref{fig:ring-atlv} (a,b)). 
The resonance frequency mode at $q_J=0$ 
becomes eventually gapless 
at $H=H_c$, being consistent with the classical spin 
configuration which starts to acquire finite in-plane 
components forming a circular vortex for $H<H_c$. 
For $H>H_c$, 
these excitations become gapped  again with the minimum 
being at $q_J=0$. For $H\gg H_c$,  a `band center'  
of these resonance frequency levels converges to  
usual ferromagnetic resonance (FMR) mode. 
Note also that two time-reversal-pair modes, 
$q_J$ and $-q_J$, are degenerate at the zero field, 
while they are not under a finite field. For the out-of-plane 
field along $+z$ direction, $\varepsilon_{q_J}<\varepsilon_{-q_J}$ 
for $q_J\gtrsim 0$ (see Fig.~\ref{fig:ring-atlv}(b,c,d)).

When a circular ring embedded into the square lattice, 
the quantum number for the atomic orbital   
reduces to either one of the following four,  
$q_J=0,\pm\frac{2\pi}{M},\frac{4\pi}{M}$. Namely, 
each ring feels an anisotropic demagnetization 
field from its surrounding rings, which 
respects four-fold rotational symmetry. 
This mixes any two states whose $q_J$ 
differs by $\frac{8\pi}{M}$ (Fig.~\ref{fig:ring-atlv} (c,d)). 
Under the four-fold rotation, these four atomic orbital 
wave functions acquire $+1$, $\pm i$ and $-1$, which 
suggests that they are essentially $s$-wave, 
$p_{\pm}=p_x\pm ip_y$-wave and $d_{x^2-y^2}$-wave 
function respectively; 
\begin{align}
\psi^{(n)}_{s}\big(\theta_j+\frac{\pi}{2}\big) &= 
\psi^{(n)}_{s}\big(\theta_j\big), \label{s-w} \\ 
\psi^{(n)}_{p_{\pm}}\big(\theta_j+\frac{\pi}{2}\big) &= \pm i \!\ 
\psi^{(n)}_{p_{\pm}}\big(\theta_j\big), \label{p-w} \\ 
\psi^{(n)}_{d_{x^2-y^2}}\big(\theta_j+\frac{\pi}{2}\big) &= 
- \psi^{(n)}_{d_{x^2-y^2}}\big(\theta_j\big). \label{d-w}
\end{align}
Every fourth levels from below are grouped together in the 
frequency space, forming a branch specified by the 
superscript index $n$ (Fig.~\ref{fig:ring-atlv}(d)); every branch 
includes the four types of wave functions,  
$s$, $p_{\pm}$, $d_{x^2-y^2}$-wave functions. 
%($n=1,\cdots,M/4$).  
The corresponding atomic orbital levels 
are arranged in the frequency space as 
\begin{align}
&\varepsilon^{(1)}_s < \varepsilon^{(1)}_{p_+} 
< \varepsilon^{(1)}_{p_-} 
< \varepsilon^{(1)}_{d_{x^2-y^2}} < \nn \\
&\ \ \varepsilon^{(2)}_{d_{x^2-y^2}} < 
\varepsilon^{(2)}_{p_{-}} 
< \varepsilon^{(2)}_{p_+} < \varepsilon^{(2)}_{s} < 
\varepsilon^{(3)}_{s} 
< \varepsilon^{(3)}_{p_+} < \cdots. \label{fre}
\end{align}

When inter-ring `exchange' processes via magnetic 
dipole-dipole interaction are included, 
these atomic orbitals constitute extended  
volume-mode bands. When neighboring branches are sufficiently 
separated from each other by the anisotropic 
demagnetization field, the volume-mode bands can be 
constructed out of each branch separately; 
\begin{align} 
&\big\{\psi^{(2m+1)}_{s},\psi^{(2m+1)}_{p_+},\psi^{(2m+1)}_{p_-}, 
\psi^{(2m+1)}_{d_{x^2-y^2}}\big\}, \label{s1} 
\end{align}
or 
\begin{align}
&\big\{\psi^{(2m+2)}_{d_{x^2-y^2}},\psi^{(2m+2)}_{p_-},
\psi^{(2m+2)}_{p_{+}}, \psi^{(2m+2)}_{s}\big\}. \label{s2}
\end{align} 
Each branch provides four volume-mode bands. A 
qualitative feature of the four volume-mode bands 
can be roughly captured by a 
two-orbital model made out of the lower two atomic orbital 
wave functions within each branch;  
\begin{align}
\big\{ \psi^{(2m+1)}_s, \psi^{(2m+1)}_{p_+} \big\}, \ \ 
{\rm or} \ \ \big\{ \psi^{(2m+2)}_{d_{x^2-y^2}}, 
\psi^{(2m+2)}_{p_-} \big\}. \label{s3}
\end{align}
This is because lower two atomic orbitals within 
each branch have less nodes than the other two 
along the ring. The inter-ring transfer 
integrals among such two are expected 
to be larger than those otherwise. 

From the symmetry point of view, a nearest 
neighbor tight-binding model composed of the lower 
two orbitals is given by; 
\begin{align}
\hat{H}_{01} &= \sum_{\bm b} \big(\varepsilon_0 \gamma^{\dagger}_{0,{\bm b}} 
\gamma_{0,{\bm b}} + \varepsilon_1\gamma^{\dagger}_{1,{\bm b}} 
\gamma_{1,{\bm b}} \big) \nn \\
&\hspace{-0.4cm} - \sum_{\bm b} 
\sum_{\mu=x,y} \sum_{\sigma=\pm} \big(a_{00} 
\gamma^{\dagger}_{0,{\bm b}} 
\gamma_{0,{\bm b}+\sigma{\bm e}_{\mu}} 
- a_{11} \gamma^{\dagger}_{1,{\bm b}} 
\gamma_{1,{\bm b}+\sigma{\bm e}_{\mu}} \big) \nn \\
& \hspace{0.5cm} - \sum_{\bm b} 
\sum_{\sigma=\pm} \big(- \sigma b_{01} \gamma^{\dagger}_{0,{\bm b}} 
\gamma_{1,{\bm b}+\sigma{\bm e}_x} + {\rm H.c.} \big) \nn \\
& \hspace{0.5cm} - \sum_{\bm b} 
\sum_{\sigma=\pm} \big(-i\sigma b_{01} \gamma^{\dagger}_{0,{\bm b}} 
\gamma_{1,{\bm b}+\sigma{\bm e}_y} + {\rm H.c.} \big). \label{sp-model}
\end{align}
Here $\gamma^{\dagger}_{0,{\bm b}}$ 
($\gamma_{0,{\bm b}}$) and 
$\gamma^{\dagger}_{1,{\bm b}}$ ($\gamma_{1,{\bm b}}$) 
stand for creation (annihilation) operators for 
parity-even and parity-odd 
atomic orbitals respectively. The subscript ${\bm b}$ 
denotes a coordinate of a center of a ring which the orbitals 
belong to. ${\bm e}_{\mu}$ 
is the primitive translation vector 
of the square lattice ($\mu=x,y$). 
The parity-even atomic orbital refers to $s$-wave or 
$d_{x^2-y^2}$-wave, while the parity-odd atomic orbital refers to 
$p_{\pm}$-wave:
\begin{align}
\big\{\varepsilon_0,\varepsilon_1\big\} 
 =\big\{\varepsilon^{(2m+1)}_{s}, \varepsilon^{(2m+1)}_{p_+}\big\}, \ \ 
{\rm or} \ 
\big\{\varepsilon^{(2m+2)}_{d_{x^2-y^2}},
\varepsilon^{(2m+2)}_{p_{-}}\big\}, \nn
\end{align}
so that $\varepsilon_0<\varepsilon_1$. 
A general observation of orbital shapes 
suggests that $a_{00}$, $a_{11}$ and $b_{01}$ are all positive 
real values under a proper gauge choice.
  
The tight binding Hamiltonian in the momentum space 
is expanded in term of the Pauli matrices as, 
$H({\bm k}) = c({\bm k}) \!\ {\bm \sigma}_0  
+ \sum^{3}_{j=1} {\bm h}_{j}({\bm k}) \!\ {\bm \sigma}_j$ 
with $h_3({\bm k}) \equiv \epsilon_0-\epsilon_1 - 
2(a_{00}+a_{11})(\cos k_x+\cos k_y)$, $h_1({\bm k})\equiv 
2 b_{01} \sin k_y$ and $h_{2}({\bm k}) \equiv 2b_{01} \sin k_x$.
In terms of a vector field ${\bm h}({\bm k})$, 
the topological Chern integer for the two volume-mode bands 
obtained from this Hamiltonian can be defined as a 
wrapping number of a normalized vector 
$\overline{\bm h}({\bm k})\equiv 
{\bm h}({\bm k})/|{\bm h}({\bm k})|$.~\cite{Volovik,Yakovenko,QWZ} 
The integer counts how many times the normalized vector wraps 
the unit sphere, when the momentum ${\bm k}$ wraps 
around the two-dimensional Brillouin zone with the torus 
geometry;~\cite{Volovik,Yakovenko,QWZ}
\begin{eqnarray}
c_{+} = -c_{-} = \int_{[-\pi,\pi]^2} \frac{d^2{\bm k}}{4\pi} \!\ \!\ 
\overline{\bm h}({\bm k}) \cdot \big(\partial_{k_x} 
\overline{\bm h}({\bm k}) 
\times \partial_{k_y} \overline{\bm h}({\bm k})\big). \nn
\end{eqnarray} 
Within a two-band model, the integer for the upper 
band ($c_+$) always has an opposite sign to that for 
the lower band ($c_-$).  
When two nearest neighboring rings are 
spatially proximate to each other, larger exchange 
integrals realize 
$\varepsilon_1-\varepsilon_0 < 4 (a_{00}+a_{11})$, 
which makes the wrapping number to be unit. 
Namely, the unit vector points at the south pole/north pole  
($\bar{\bm h}=(0,0,-1)$/$(0,0,+1)$) 
at ${\bm k}=(0,0)$/$(\pi,\pi)$, while the vector winds 
once around the south pole/north pole   
when ${\bm k}$ rotates once 
around the ${\bm k}=(0,0)$/$(\pi,\pi)$. This observation 
suggests that the Chern integers for two bands 
obtained from eq.~(\ref{sp-model})  
become $\{c_-,c_+\}=\{-1,+1\}$. When the out-of-field 
direction is reversed, $p_{+}$ and $p_{-}$ are exchanged in 
\begin{figure}[ht]
   \includegraphics[width=85mm]{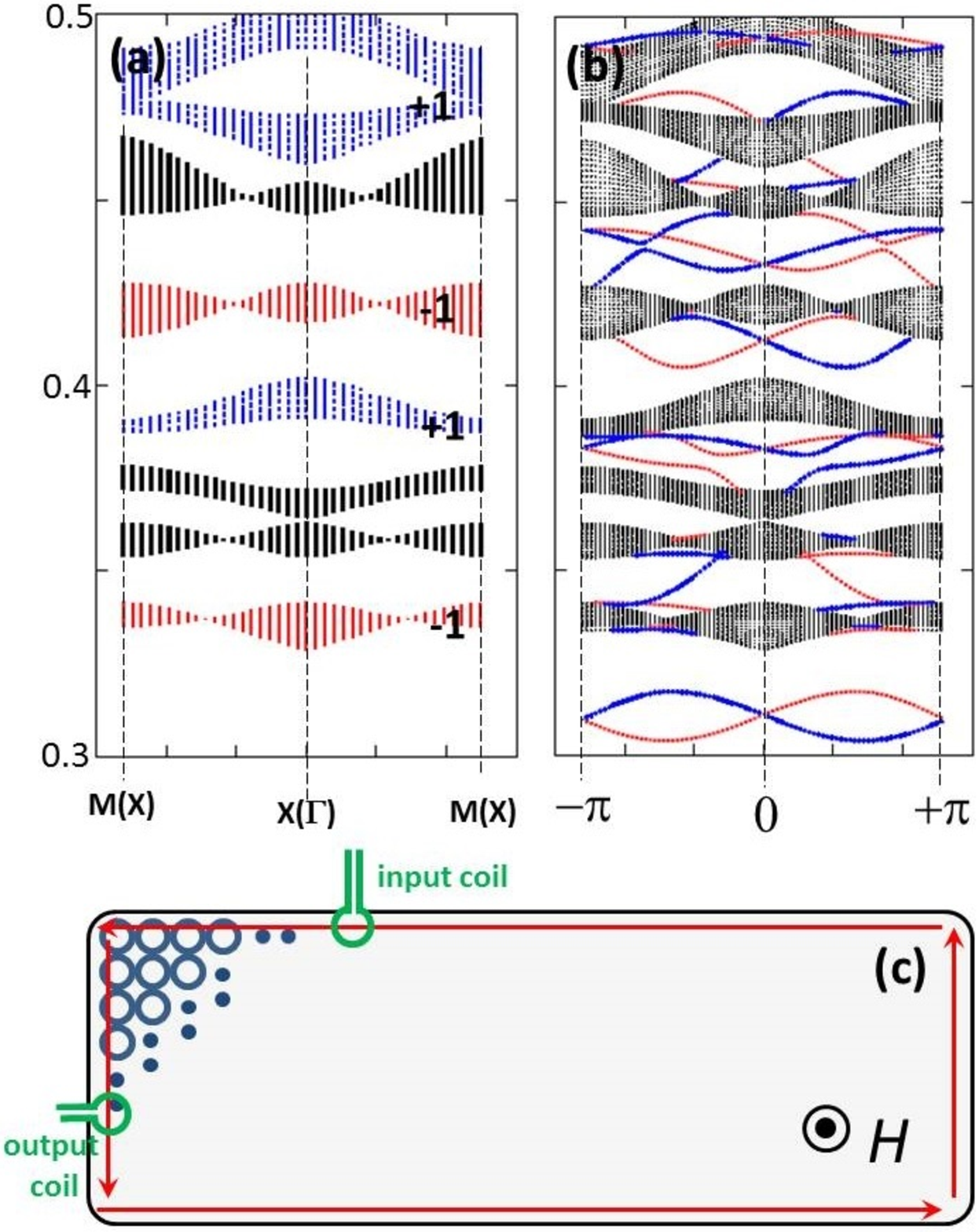}
\caption{ (Color online) Wavelength-frequency 
dispersions 
for spin-wave excitations for $H_d < H< H_c$ $(H=0.94H_c)$.  
(a) A side-view of lowest 8 volume-mode bands with the Chern 
integer. The red bands have $-1$ Chern integer, 
while blue bands have $+1$. The dispersions  
are calculated with periodic boundary conditions for 
both $x$ and $y$-directions. Since the 7th 
and 8th lowest band have frequency degeneracies 
around $M$-points, only the sum of their integers 
is quantized to $+1$. (b) Spin-wave excitations 
calculated with an open/periodic boundary condition 
along the $y$/$x$-direction respectively. 
The resonance frequencies are given as a 
function of the wave vector along the 
$x$-direction. The system along the $y$-direction 
includes 18 square-lattice unit cells ($L=18$). More than 
$75\%$ of amplitudes of eigen wave functions with red points 
are localized within $y=1$ and $y=2$, 
while those with blue points are localized 
within $y=L-1$ and $y=L$ (edge modes). 
Compared with Fig.~(a), the calculated 
spectra have additional 
spin-wave modes which are localized 
along the edges. 
(c) With the out-of-plane field up-headed, the chiral 
edge modes rotate in the counterclockwise way.}
\label{fig:1.2-1}
\end{figure}
Fig.~\ref{fig:ring-atlv}(d) and eqs.~(\ref{s1},\ref{s2}), 
which changes the sign of 
the last term in eq.~(\ref{sp-model}) and that of 
$c_{\pm}$. Note also that, to have the   
non-zero wrapping number, it is essential that 
`wave function character' for the lower/higher  
band at ${\bm k}=(\pi,\pi)$ is 
parity odd/even atomic orbital, while 
that at the ${\bm k}=(0,0)$ is 
parity-even/odd one (`band inversion').~\cite{BTZ,FK} When 
$\varepsilon_1-\epsilon_0 > 4 (a_{00}+a_{11})$, 
 wave function character of the lower (higher) band 
at ${\bm k}=(\pi,\pi)$ and that of ${\bm k}=(0,0)$ 
have same parity, so that the unit vector always 
stays within the southern hemisphere, 
irrespective of the momentum ${\bm k}$; 
the wrapping number always reduces to zero.

The argument so far suggests that, in the presence 
of larger inter-ring transfer integrals, 
the distribution of the Chern integers for 
soft volume-mode bands at $H_d<H<H_c$ 
can be non-trivial and is composed of a 
sequence of $\{-1,+1,0,0\}$ from below; 
\begin{align}
&\big\{c_1,c_2,c_3,c_4\!\ \big|,c_5,c_6,c_7,c_8\!\ \big|,\cdots\big\} \nn \\
& \ \ = \big\{-1,+1,0,0\!\ \big|,-1,+1,0,0\!\ \big|,\cdots\big\} \label{6}
\end{align} 
where $c_n$ denotes the integer for the $n$-th lowest 
band (see also appendix A for general definition of the 
topological Chern integer for volume-mode spin-wave bands). 
An explicit calculation of the Chern integers for volume-mode 
bands within $H_d < H< H_c$ based on a linearized 
Landau-Lifshitz equation confirms this feature with a 
minor modification. In the actual 
calculation, we also observed that, within each branch,  
another band inversion is often 
induced by relatively stronger exchange integrals  
between higher two atomic orbitals and the 2nd lowest atomic 
orbital, which transfer the non-zero integer of 
the 2nd lowest band into the 3rd or 4th lowest bands in each 
branch, $\{-1,+1,0,0\} \rightarrow   
\{-1,0,+1,0\}$ or $\{-1,0,0,+1\}$. Which comes true 
among these three, i.e.  
$\{-1,+1,0,0\}$, $\{-1,0,+1,0\}$ and $\{-1,0,0,+1\}$, 
depends on specific branch and other details, while 
the integer for the lowest band ($-1$) remains 
intact in every branch ,e.g.  %This makes  
%the distribution of the Chern integers for whole volume-mode 
%bands something like, e.g.    
\begin{align}
&\big\{c_1,c_2,c_3,c_4\!\ \big|,c_5,c_6,c_7,c_8\!\ \big|,\cdots\big\} \nn \\
& \ \ = \big\{-1,0,0,+1\!\ \big|,-1,0,\alpha,1-\alpha \!\ \big|,\cdots\big\},  
\label{7}
\end{align}
(Fig.~\ref{fig:1.2-1}(a)). 

General arguments~\cite{SO1,SO2} 
based on a bulk-edge correspondence in 
IQH physics~\cite{TKKN,Hal,Hat} 
dictate that the Chern 
integers for the volume-mode bands shown in eq.~(\ref{7}) 
lead to counterclockwise rotating spin-wave edge 
modes, whose chiral dispersion connects in the frequency 
space a volume-mode band with $-1$ Chern integer 
and that with $+1$ Chern integer. In fact, the existence 
of such chiral edge modes are confirmed by 
quantitative band calculations based on a linearized 
Landau-Lifshitz equation with open boundary condition  
(Fig.~\ref{fig:1.2-1}(b)). 
Again, reversing the out-of-field direction ($+z \rightarrow - z$) 
results in the sign change of $c_n$, 
which changes the chiral 
direction of the edge modes from counterclockwise to 
clockwise (Fig.~\ref{fig:1.2-1}(c)).

When the out-of-plane field is less than the `critical' field, 
$H<H_d$, the lower spin-wave volume mode bands are 
from those atomic orbitals having higher total 
angular momentum $q_J = \pi$. Compared to those around 
$q_J=0$, such orbitals have many 
nodes along the ring; their wave functions change sign under 
the translation only by one spin, 
e.g. 
\begin{eqnarray}
\psi_{q_J=\pi}(\theta_{j+1})=-\psi_{q_J=\pi}(\theta_j). \nn
\end{eqnarray}
Due to this many-node structure, 
transfer integrals between the higher angular momentum 
orbitals ($q_J\simeq \pi$) become much smaller 
than those between orbitals with lower angular 
momentum ($q_J \simeq 0$). 
As a result, low-frequency volume-mode bands 
for $H<H_{d}$ have tiny dispersions, which  
can hardly fulfill the band inversion condition,  
$|\epsilon_{1}-\epsilon_{0}| < 4 (a_{00}+a_{11})$;
we thus cannot expect the chiral spin-wave 
edge modes.

\begin{figure}[th]
   \includegraphics[width=80mm]{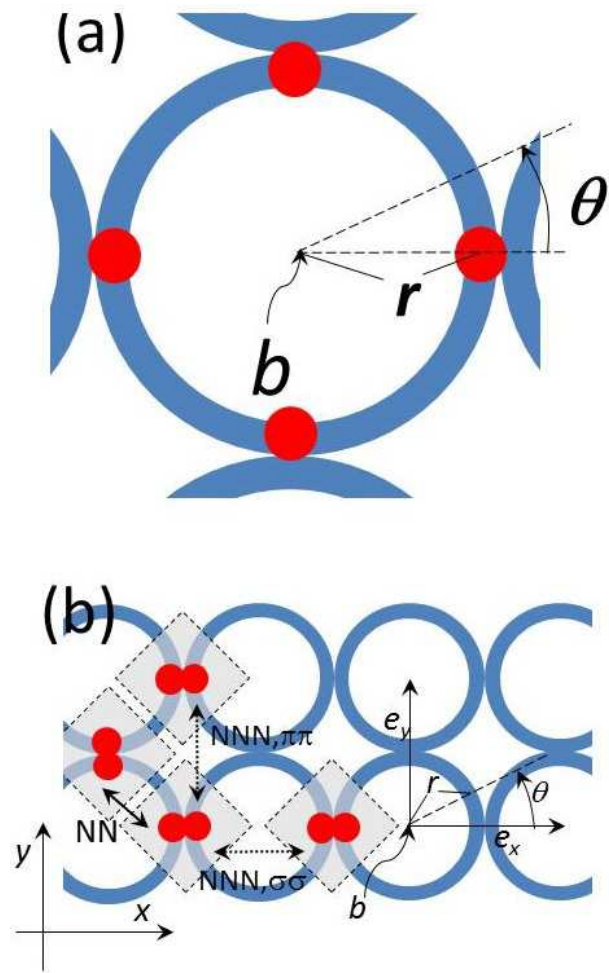}
\caption{ (Color online) (a) Four corners  
in a ring (red regions;  
$\theta=0,\frac{\pi}{2},\pi,\frac{3\pi}{2}$) feel 
larger demagnetization field than 
other regions. ${\bm b}$ denotes a 
coordinate of the center of the ring, while
$r$ is a radius of the ring. 
(b) two-orbital tight-binding model 
with nearest-neighbor (`NN' in the figure) 
inter-cluster transfer integral  
(${\bm H}_0$), next-nearest-neighbor $\sigma$-$\sigma$ 
coupling (`NNN,$\sigma\sigma$') inter-cluster transfer integral 
(${\bm H}_{1}$ with $c$) and 
next-nearest-neighbor $\pi$-$\pi$ 
coupling (`NNN,$\pi\pi$') inter-cluster transfer integral 
(${\bm H}_{1}$ with $c'$). The in-phase orbital 
(red peanut-shape item) at the $x$-link 
is extended along the $x$-direction, while 
that at the $y$-link is along the $y$-direction. 
${\bm e}_x$ and ${\bm e}_y$ denote the primitive 
translation vectors of the square lattice. } 
\label{fig:2-orbital-TB}
\end{figure}

Above the saturation field ($H>H_c$), 
the four-fold rotational anisotropy in 
the demagnetization field becomes stronger. 
When the classical spin configuration 
becomes fully polarized along the out-of-plane field, 
spins in a ring which are proximate to its four 
nearest neighboring rings especially feel stronger 
demagnetization fields than those spins in the ring 
which are not. In terms of the angle variable $\theta$ 
defined as 
${\bm r} \equiv {\bm b} + r(\cos\theta,\sin\theta)$ 
(${\bm b}$ denotes a coordinate of a center of 
the ring at which a spin at ${\bm r}$ is included 
and $r$ is the radius of the ring; 
see fig.~\ref{fig:2-orbital-TB}(a)), 
these spins are at the four corners of a ring,  
$\theta=0,\frac{\pi}{2},\pi,\frac{3\pi}{2}$ 
respectively. 
As a result of this strongly anisotropic demagnetization 
field, soft spin-wave 
excitations for $H>H_c$ are highly 
localized around these four corners.

From this point of view, we 
made another tight binding model 
for soft spin-wave bands, which is valid only 
above the saturation field (appendix B). Thereby, 
we first took into account proximate `exchange 
process' which transfers a spin in a corner of  a ring 
into its closest corner of the nearest 
neighboring ring. The inclusion of such  
exchange process leads to in-phase and 
out-of-phase orbital wave functions formed by 
these two spins. These `atomic-orbital'  
wave functions are on a 
center of a link connecting two nearest neighboring 
rings (red peanut-shape items in Fig.~\ref{fig:2-orbital-TB}(b)). 
It turns out that, when the field 
is not too close to the saturation field, the in-phase 
atomic orbital level becomes lower than the out-of-phase 
orbital level (see Appendix B for the argument). 

The square lattice has two inequivalent  
links within its unit cell, the link along the 
$x$-axis (`$x$-link') and that along the 
$y$-axis (`$y$-link'). Each link provides in-phase  
and out-of-phase orbital wave functions. Since the  
out-of-phase wave function has a node at the center, 
while the in-phase one does not, inter-link transfer integrals 
between out-of-phase orbitals becomes smaller 
than those between in-phase orbitals. Being 
interested in spin-wave bands with larger band width, 
we focus only on the in-phase orbital wave functions. 

A transfer integral between $x$-link and its 
nearest neighbor $y$-link becomes complex-valued;
\begin{align}
H_{0} &= \sum_{{\bm b}} \Big\{ (i a + b) \!\ 
\beta^{\dagger}_{{\bm b}+\frac{{\bm e}_y}{2}} 
\beta_{{\bm b}+\frac{{\bm e}_x}{2}} 
\nn \\
&\ \ \ - (i a+b) \!\ 
\beta^{\dagger}_{{\bm b}+{\bm e}_y+\frac{{\bm e}_x}{2}} 
\beta_{{\bm b}+\frac{{\bm e}_y}{2}} \nn \\
& \ \ \ \ + (i a + b) \!\ 
\beta^{\dagger}_{{\bm b}+{\bm e}_x + \frac{{\bm e}_y}{2}} 
\beta_{{\bm b}+{\bm e}_y+\frac{{\bm e}_x}{2}} \nn \\
& \ \ \ \ \ \ - (i a+ b) \!\ 
\beta^{\dagger}_{{\bm b}+\frac{{\bm e}_x}{2}} 
\beta_{{\bm b}+{\bm e}_x+\frac{{\bm e}_y}{2}} + {\rm h.c.} \Big\} 
\label{h0}
\end{align}  
with real-valued $a$ and $b$.   
$\beta_{{\bm b}+\frac{{\bm e}_{x}}{2}}$ and 
$\beta_{{\bm b}+\frac{{\bm e}_{y}}{2}}$ represent 
annihilation operators for the in-phase orbital on 
the $x$-link (whose center is at 
${\bm b}+\frac{{\bm e}_{x}}{2}$) and 
that on $y$-link (at 
${\bm b}+\frac{{\bm e}_{y}}{2}$) respectively. 
`$ia\equiv a e^{i\theta}$' 
with $\theta=\frac{\pi}{2}$ in eq.~(\ref{h0}) 
comes from 90$^{\circ}$ degree angle 
subtended by the two nearest neighbor 
orbitals at ${\bm b}+\frac{{\bm e}_x}{2}$ and 
at ${\bm b}+\frac{{\bm e}_y}{2}$   
and a center of the ring at ${\bm b}$. `$b$' in eq.~(\ref{h0}) 
results from a finite particle-hole mixing (see appendix B 
for the derivation of eq.~(\ref{h0})).    
A band structure obtained from $H_0$ has  
two frequency bands which form gapless Dirac cone spectra  
at ${\bm k}=(\pi,0)$ and $(0,\pi)$.

\begin{figure}[ht]
   \includegraphics[width=85mm]{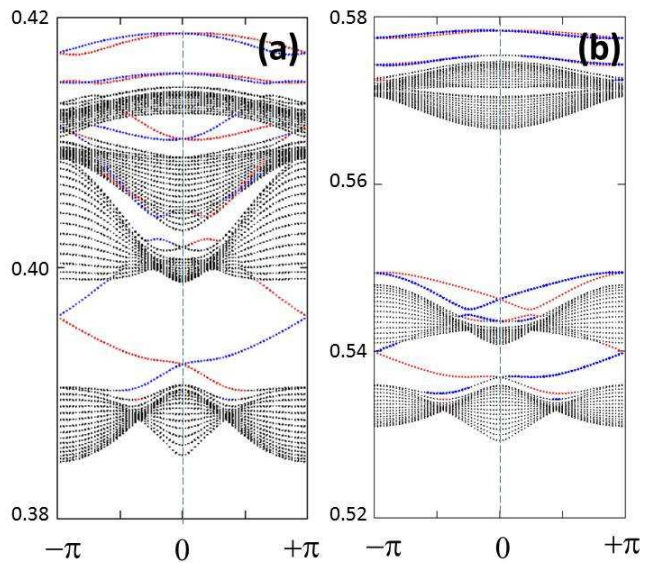}
\caption{ (Color online) Wavelength-frequency 
dispersions for lowest four volume-mode bands 
and chiral edge modes in $H>H_c$ 
(a) $H =1.09 H_c$ (b) $H=1.17 H_c$. The dispersion 
are obtained with an open/periodic boundary condition 
along the $y$/$x$-direction, 
where the resonance frequencies for spin wave excitations 
are given as a function of the wave vector along the 
$x$-direction. We used the same system size 
along the $y$-direction as in Fig.~\ref{fig:1.2-1} 
and the same definition of red and blue points  
as in Fig.~\ref{fig:1.2-1}. In both (a) and (b), the lowest 
two volume modes (black points) 
consist of the in-phase atomic 
orbitals on $x$-link and $y$-link, while the upper two 
volume modes mainly consist of out-of-phase orbitals 
on these two links. The spectra clearly contain  
a chiral edge mode connecting the lowest two 
volume-mode bands. Compared to 
the lowest two bands, the 3rd and 4th lowest bands have 
smaller band width and no band gap in between. This is 
because, contrary to the in-phase atomic orbital, 
the out-of-phase atomic orbital has a node at the center 
of each link, which results in smaller transfer integrals. 
Compared (a) and (b), note also that a frequency 
spacing between the in-phase atomic orbital level and 
the out-of-phase atomic orbital level increases 
on increasing the field (see appendix B for 
the reasoning). }
\label{fig:1.5-2}
\end{figure}

A finite transfer between 
the nearest $x$-links and that between the 
nearest $y$-links endows the gapless Dirac cone spectra 
with a finite mass. The transfer takes a form of,
\begin{align}
H_1 &= \sum_{{\bm b}} \Big\{c \!\ 
\beta^{\dagger}_{{\bm b}+\frac{{\bm e}_x}{2}} 
\beta_{{\bm b} - \frac{{\bm e}_x}{2}} 
+ c \!\ \beta^{\dagger}_{{\bm b}+\frac{{\bm e}_y}{2}} 
\beta_{{\bm b} - \frac{{\bm e}_y}{2}} \nn \\
& \ \ \  + c^{\prime} \!\ 
\beta^{\dagger}_{{\bm b}+{\bm e}_y +\frac{{\bm e}_x}{2}} 
\beta_{{\bm b} + \frac{{\bm e}_x}{2}} 
+ c^{\prime} \!\ 
\beta^{\dagger}_{{\bm b}+{\bm e}_x 
+\frac{{\bm e}_y}{2}} 
\beta_{{\bm b} + \frac{{\bm e}_y}{2}} + {\rm h.c.}
 \Big\}, \label{h1}  
\end{align}
with real-valued $c$ and $c'$. Now that orbital wave function 
at the $\mu$-link is extended along the $\mu$-axis, 
`$c$' stands for the ($\sigma,\sigma$)-coupling next nearest 
neighbor (NNN) transfer integral, while `$c'$' stands 
for the $(\pi,\pi)$-coupling 
NNN transfer integral (Fig.~\ref{fig:2-orbital-TB}(b)). 
Amplitudes of transfer integrals are  
inversely proportional to the cubic in 
distance, so that $|c|>|c'|$. A finite $|c-c'|$ induces 
a gap in the gapless Dirac cone spectra.

The Chern integers for these two spin-wave 
bands can be evaluated from 
the wrapping number of the normalized vector 
$\overline{\bm h}({\bm k})\equiv {\bm h}({\bm k})/|{\bm h}({\bm k})|$. 
For eqs.~(\ref{h0},\ref{h1}), 
$h_1({\bm k})=4b \sin\frac{k_x}{2} \sin \frac{k_y}{2}$, 
$h_2 ({\bm k})= 4a  \cos \frac{k_x}{2} \cos \frac{k_y}{2}$ and 
$h_3({\bm k})=2(c-c^{\prime})(\cos k_x -\cos k_y)$. When 
the momentum rotates around ${\bm k}=(\pi,0)$ / 
$(0,\pi)$, $\overline{\bm h}({\bm k})$ rotates around the 
south pole/ north pole once for $c>c'$; the winding 
numbers are $\pm 1$. More generally, 
the integers for these two bands are 
$\{c_{-},c_{+}\}=\{+1,-1\}$ from below 
for $(c-c^{\prime})\cdot a \cdot b >0$, 
while $\{-1,+1\}$ for $(c-c^{\prime})\cdot a \cdot b < 0$.  
In either case, there appears a chiral edge 
mode within the band gap, whose sense of 
rotation along the boundary is clockwise 
for the former case, while counterclockwise for the latter case. 
A primitive evaluation suggests that 
$a >0$, $b >0$ and $c<c^{\prime}<0$ (appendix B), 
so that a counterclockwise chiral edge mode is expected. 
In fact, the counterclockwise chiral edge mode 
is observed within a band gap between the lowest 
and the 2nd lowest volume-mode band 
for a wide field range of $H>H_c$ (Fig.~\ref{fig:1.5-2}). 
Contrary to the effective $s$-$p_{\pm}$ 
model for $H_d <H<H_c$, the band gap and the chiral 
edge mode in the present two-orbital model 
persist for a wider range of $H>H_c$. This is because 
any symmetries in the model requires 
neither $c = c^{\prime}$ nor $a=0$, 
while $b$ vanishes only in the large $H$ 
limit (appendix B). This feature is indeed justified 
by the micromagnetic simulation in sec.~IV. 

\section{disk model}
\begin{figure}[t]
   \includegraphics[width=85mm]{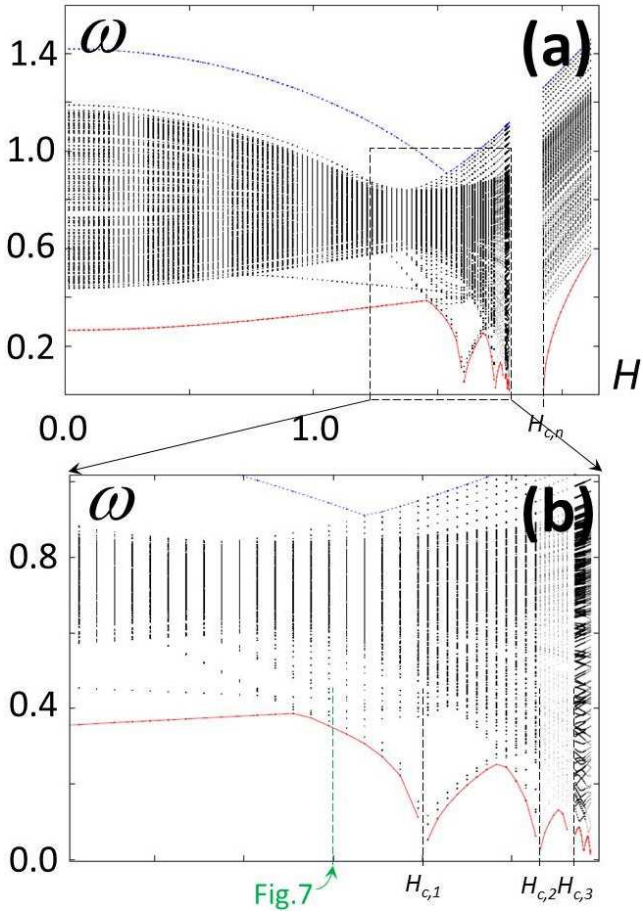}
\caption{ (Color online) (a,b) Distribution of resonance 
frequency levels of magnon modes in a single circular 
magnetic disk as a function of the out-of-plane field.  
The four-fold-rotational demagnetization field from 
other disks are also included in the calculation. 
The red curve plots 
the lowest resonance frequency level as a function 
of the out-of-plane field, while the blue curve  
represents the highest resonance frequency level. 
We simulate the circular magnetic disk by a cluster of spins,  
respecting the circular symmetry as much as possible. 
For a given radius $R$, we discretize $R$ into $n$ pieces. 
For the radial coordinate ranging from $[\frac{Rj}{n},\frac{R(j+1)}{n}]$ 
($j=0,1,\cdots,n-1$), we discretize the azimuth coordinate 
into $4(2j+1)$ pieces, so that area of each element is 
same, $\frac{\pi R^2}{4 n^2}$. We put a spin at the 
center of each element specified  by 
$(x,y)=r_{j} (\cos \theta_{j,m},\sin\theta_{j,m})$ 
with $r_j=\frac{R(2j+1)}{2n}$ and 
$\theta_{j,m}=\frac{\pi (2m+1)}{4(2j+1)}$ ($m=0,\cdots,4(2j-1)$). 
In the calculation, we take $n=8$, so that a cluster has 
$256$ spins.}
\label{fig:cir-atlv}
\end{figure}

Let us next consider spin-wave excitations in 
circular disk model. We simulate the magnetic disk by  
a cluster of many spins, each of which has a same 
volume element. The spins are distributed  as 
homogeneously in space as possible (see the caption 
of Fig.~\ref{fig:cir-atlv}). Physically, a linear 
dimension of the volume element should be on 
the order of short-ranged exchange interaction 
length $l_{\rm ex}$. 
The spins are coupled with one another via magnetic 
dipole-dipole interaction.

A circular vortex structure minimizes the 
magnetostatic energy of the disk at the zero 
field, while the field induces a finite out-of-plane 
magnetization. Suppose that spins are 
nearly polarized along the field, while any of them  
are not yet fully polarized. Being surrounded by 
many others, spins around the center of 
a disk feel the strongest demagnetization 
field, while the demagnetization field around the 
boundary is smallest. Thus, spins at   
the boundary become fully polarized first by a relatively 
lower field, $H_{c,1}$, while spins around the center 
become fully polarized at last by a 
relatively higher field, $H_{c,n}(>H_{c,1})$.  
In the present discrete spin model, 
these two critical fields encompasses a couple of 
other critical fields ($H_{c,1}<H_{c,2}<H_{c,3}<\cdots<H_{c,n}$), 
at which interior spins get fully polarized successively 
from the outer to the inner on increasing the field.

\begin{figure}[t]
   \includegraphics[width=85mm]{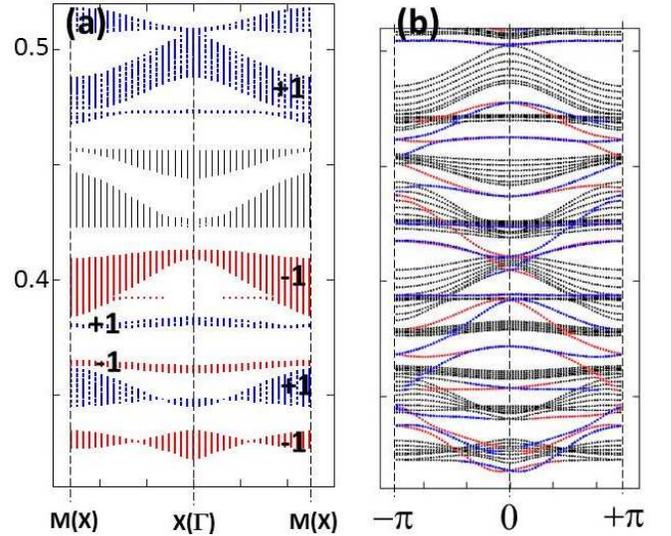}
\caption{ (Color online) Wavelength-frequency 
dispersions for spin-wave excitations for 
$H < H_{c,1}$ (see Fig.~\ref{fig:cir-atlv}(b)) 
(a) wavelength-frequency dispersion 
for volume-mode bands with the Chern integer 
($H=0.94H_{c,1})$. The dispersions are calculated with periodic boundary 
conditions for both $x$ and $y$-directions.
(b) wavelength-frequency dispersion 
for volume-mode bands and edge-mode bands 
($H=0.94H_{c,1}$) calculated with an open/periodic 
boundary condition along the $y$/$x$-direction. 
Resonance frequencies are given 
as a function of the wave vector along the 
$x$-direction. The system along the $y$-direction 
includes 9 unit cells ($L=9$). More than $50\%$ of 
amplitudes of eigen 
wave functions with red points are localized only 
within $y=1$, while those for blue points are localized 
from $y=L$ (edge modes). 
Compared with Fig.~(a), the spectra have additional 
spin-wave modes which are localized 
along the edges, 
whose chiral dispersion connect spin-wave volume modes 
bands with opposite Chern integers }
\label{fig:cir-band}
\end{figure}

Correspondingly, spin wave excitations, which 
are fully gapped at $H=0$, become gapless 
or significantly softened at each of these 
critical fields, $H=H_{c,1},H_{c,2},\cdots$ (Fig.~\ref{fig:cir-atlv}). 
Especially, the soft magnons around $H=H_{c,1}$ are localized 
around the boundary of the disk, while those 
around $H=H_{c,n}$ are localized at the center. 
In a single magnetic disk, spin-wave 
excitations have the total angular momentum $q_J$ 
as a good quantum number. All the soft magnons around 
these critical fields come from $q_J=0$, so as to 
be consistent with the classical spin configuration. 
In the presence of the four-fold rotational demagnetization 
field, these soft magnons take a form of 
either $s$-wave ($n_J=0$), $p_{\pm}$-wave ($n_J=\pm1$) 
or $d_{x^2-y^2}$-wave ($n_J=2$) 
atomic orbital. As in the ring model, an 
inter-disk exchange process via the dipolar interaction makes 
these atomic orbitals to form extended volume-mode 
bands.

Since the soft magnons around $H = H_{c,1}$ are  
localized around the boundary of the disk, the inter-disk 
transfer integrals between these magnons become   
larger and soft volume-mode bands 
around $H  =  H_{c,1}$ become similar to what we 
observed in the 
ring model at $H \simeq H_{c}$; the distribution of 
Chern integers for a set of these four bands 
becomes either $\{-1,+1,0,0\}$,  $\{-1,0,+1,0\}$, or 
$\{-1,0,0,+1\}$ from below (Fig.~\ref{fig:cir-band}(a)). 
Again, this leads to a counterclockwise chiral 
edge mode between these two (Fig.~\ref{fig:cir-band}(b)).  

On the other hand, the soft magnons in $H \gtrsim H_{c,n}$ 
are localized around the center of the disk, so that the 
inter-disk transfer integrals between these atomic orbitals are 
very small. As a result, soft volume-mode bands 
in $H \gtrsim H_{c,n}$ have tiny dispersions, where we 
cannot expect any band inversion mechanism.    

\section{micromagnetic simulation}  

\begin{figure}[htp]
   \includegraphics[width=87mm]{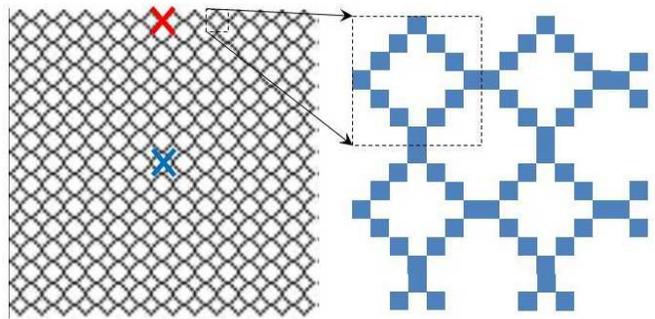}
\caption{ (Color online) magnetic superlattice 
with 14 $\times$ 14 unit cell. (right) a unit cell 
contains 12 ferromagnetic grain forming a 
square ring. Each grain is cubic-shape with its 
linear dimension $5$ nm. (left) To excite 
volume-mode/edge-mode excitations, we apply 
a pulse field at the center/boundary of the superlattice 
(blue/red crossed point) respectively.} 
\label{fig:mc3}
\end{figure}

In order to uphold the existence of the chiral edge mode  
in the proposed magnetic superlattices, we perform a 
micromagnetic simulation by solving the 
Landau-Lifshitz-Gilbert equation in terms of the 
4th order Runge-Kutta method with a unit time 
step 1ps. Fig.~\ref{fig:mc3} shows an entire magnetic 
superlattice, which contains 14 $\times$ 14 unit cells 
with open boundaries. Each unit cell contains 
12 ferromagnetic grains, forming a square-shape 
ring. Each grain is 5-nanometer cube. Note also that, not 
including any short-range exchange interaction (see below), 
the following result is scalable; provided that each ferromagnetic 
grain behaves as single spin, the size of the grain can be much 
larger than 5-nanometer and the scale of resonance 
frequency and saturation field still remain unaltered. 
The saturation magnetization and Gilbert damping 
coefficient of the ferromagnetic grain are set to 135300 A/m 
and 1.0$\times$10$^{-5}$ respectively. 
We regard each nanograin as a uniform magnet, 
assigning single spin degree of freedom to each grain. 
Different ferromagnetic nanograins are coupled with one another 
through the magnetic dipole-dipole interaction. Under 
a static out-of-plane field (along the $z$ direction) 
greater than 620 Oe, a stable spin configuration 
becomes fully polarized along the field, while the configuration 
acquires finite in-plane components below 620 Oe. We studied  
spin-wave excitations above the saturation field  
($\gtrsim$ 620 Oe).

\begin{figure}[htp]
   \includegraphics[width=87mm]{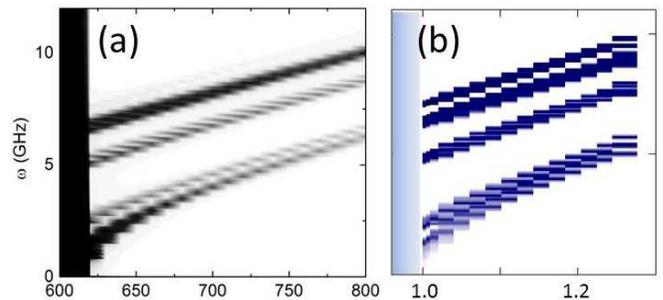}
\caption{ (Color online) (a) Contour plot of the integrated 
power specctrum $A(\omega)$ as a function of 
the static out-of-plane field $H$, where the initial pulse field 
is applied at the center of the magnetic superlattice. 
The out-of-plane field is greater than the saturation field 
($\simeq$ 620 Oe). (b) Contour plot of the density of 
state for volume-mode bands obtained from spin-wave 
calculations on the same magnetic superlattice. The horizontal 
axis is the static out-of-plane field, where the unit 
is taken to be the saturation field. In both figures, darker 
regions have higher intensities.} 
\label{fig:bulk}
\end{figure} 

To study spin wave modes in a broad frequency 
range at once, we apply a pulse magnetic 
field in a transverse ($x$) direction (pulse 
time $1$ ps and amplitude 
$1$ Oe). We then calculate 
a time evolution of magnetization 
dynamics afterward, and take a Fourier transformation 
of the transverse moments with respect to time; 
\begin{align}
s_{+}(X,Y,\omega) \equiv \sum_{j=0}^{n-1} m_{+}(X,Y,j \Delta T) 
\exp\left( 2\pi {\rm i} \omega j \Delta T \right) \label{ps-local} 
\end{align}
with $m_{+}(X,Y,t)\equiv m_x(X,Y,t)+
{\rm i}m_y(X,Y,t)$, $\Delta T=100$ ps and $n=1024$. 
An amplitude of the frequency power spectrum,  
$|s_{+}(X,Y,\omega)|$, represents a sort of local density of 
state of spin-wave modes at the resonance frequency 
$\omega$. When integrated over the two-dimensional 
space coordinates, ($X,Y$), 
the power spectrum represents the total 
density of states at $\omega$;
\begin{align}
A(\omega) \equiv \sum_{X,Y}  \big|s_{+}(X,Y,\omega)\big|, \label{ps}
\end{align} 
\begin{figure}[htp]
   \includegraphics[width=95mm]{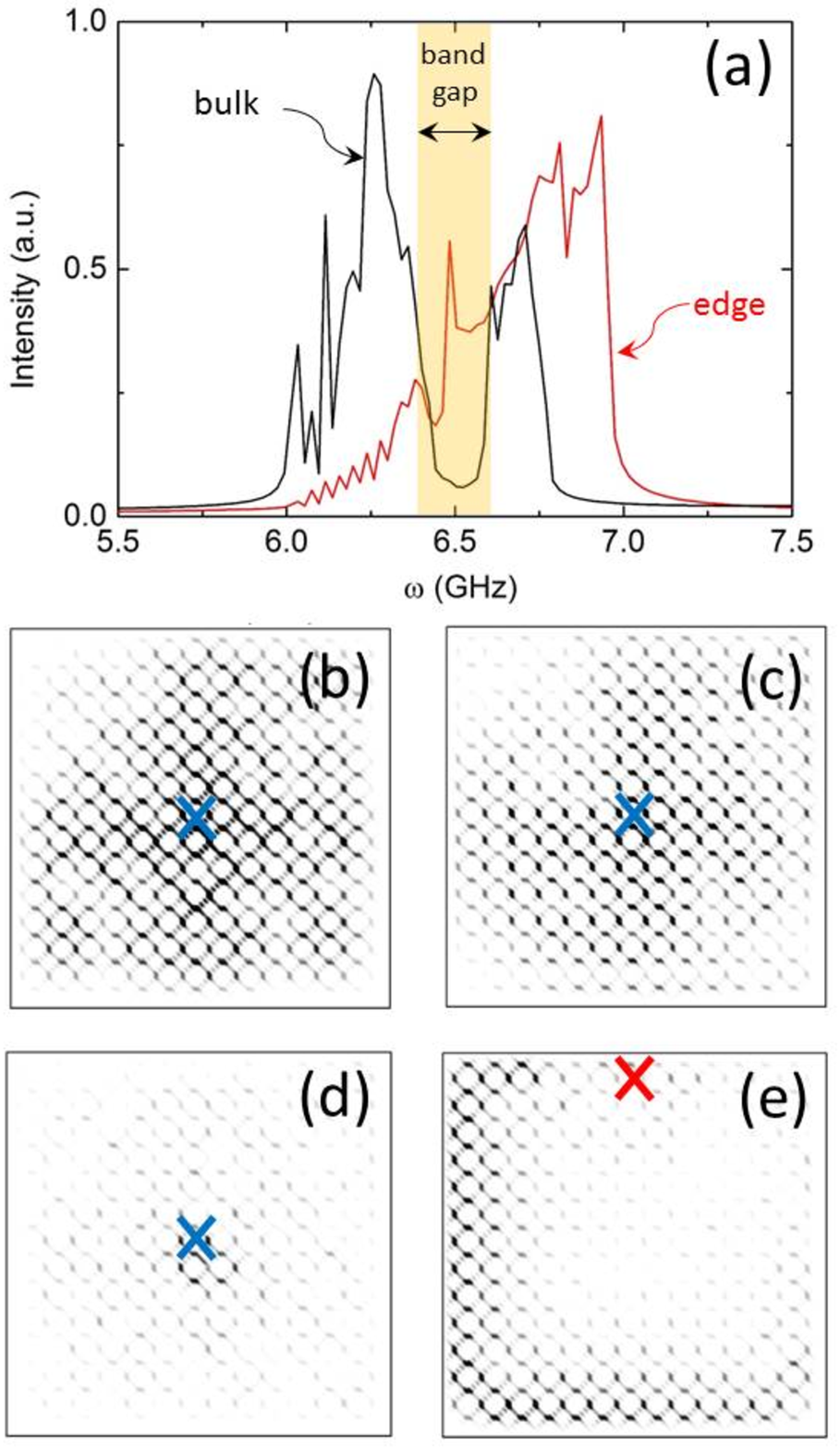}
\caption{ (Color online) (a) integrated power spectra 
calculated with the pulse field at the center (blue) 
and at the boundary (red). The static out-of-plane field 
is set to 800 Oe. (b-d) spatial-resolved  
power spectra $|s_{+}(X,Y,\omega)|$ 
calculated with the pulse field at the center (blue crossed point); 
(b) $\omega=$ 6.25 GHz, (c) 6.69 GHz, (d) 6.54 GHz. 
(e) spatial-resolved power spectrum calculated 
with the pulse field at the boundary (red crossed point) with 
$\omega=6.54$ GHz.} 
\label{fig:lw2}
\end{figure}
(see Fig.~\ref{fig:bulk} for a comparison between the integrated power 
spectra and the total density of state obtained from spin-wave 
calculations). 
For the purpose of studying volume modes and edge modes 
selectively, we did two micromagnetic simulations; one with 
the initial pulse field 
applied at the center of the system, 
exciting volume modes, and the other 
with the pulse field  applied 
near the boundary of the system, 
exciting edge modes. The power spectra obtained 
from these separate simulations are regarded as 
the density of states of volume/edge-mode bands 
respectively.

\begin{figure}[htp]
   \includegraphics[width=85mm]{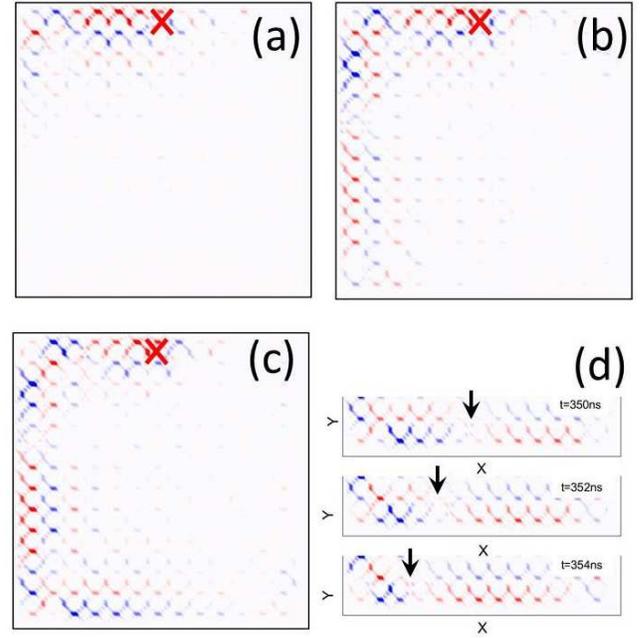}
\caption{ (Color online) snap shots of a transverse 
magnetic moment after the a.c. field is applied at $t=0$; 
(a): $t=100$ns, (b): 200ns, (c) 300ns, (d) t=350, 352, 354ns.
The frequency of the a.c. field and the static out-of-plane 
field is set to 654GHz and 800 Oe respectively. Color 
specify the sign of the transverse moment (red is for positive and 
blue is for negative). In (a-c), the spin density propagates 
in the counterclockwise direction, while, in (d), the node of the 
transverse moment (indicated by black arrows) 
moves in the clockwise direction; 
the phase velocity is opposite to the group velocity.} 
\label{fig:snap}
\end{figure}  
Fig.~\ref{fig:bulk}(a) shows a contour plot of the integrated power spectrum 
$A(\omega)$ as a function of the static out-of-plane field 
($\ge$ 620 Oe). The initial pulse field is applied at the 
center of the superlattice. On the whole, the spectrum 
composes of three major responance frequency regimes; 
for H=$800$Oe, these three are ranged over 6$\sim$7GHz, 
8.5$\sim$9GHz, and  9.5$\sim$11GHz respectively.  
Fig.~\ref{fig:bulk}(b) shows a contour plot of the density 
of states of volume-mode bands obtained from a  
spin-wave calculation on the same magnetic superlattice. Since 
the superlattice has 12 spins within each unit cell, it 
has 12 volume-mode bands. 
A comparison reveals that the first and second lowest resonance frequency 
regimes found in $A(\omega)$ includes two volume-mode 
bands respectively, while the third resonance frequency regime 
includes remaining 8 bands. A comparison with the spin-wave 
analyses also shows that the lowest two volume-mode 
bands can be well reproduced by the two-orbital 
tight-binding model introduced in eqs.~(\ref{h0},\ref{h1}); 
the lowest two bands are mainly composed of the in-phase 
orbital wavefunction localized at the nearest neighbor 
$x$-link and that of the $y$-link. Thereby, they 
are essentially same as the lowest two  
bands found in the sec.~II ($H>H_c$), and thus we expect that 
the chiral edge mode goes across a band gap between 
these two (see Fig.~\ref{fig:1.5-2}).

Fig.~\ref{fig:lw2}(a) shows the integrated power spectra 
within the lowest resonance frequency regime. 
The spectrum for 
volume-mode bands (spectrum obtained with the initial 
pulse field applied at the center of the system) 
comprises of two major humps; one 
ranges from 6.0GHz to 6.4GHz and the other 
from 6.6GHz to 6.8GHz (black line in Fig.~\ref{fig:lw2}(a)). 
They correspond to the lowest two volume-mode bands. 
In fact, the spatial-resolved power spectra within these 
two frequency regimes are extended over the system 
(Fig.~\ref{fig:lw2}(b,c)), while the system remains intact 
against those pulse fields within a band gap regime  
6.4GHz $\sim$ 6.6GHz. (Fig.~\ref{fig:lw2}(d)). 
When the pulse field is applied at the boundary 
of the system, however, the integrated spectrum 
has a significant weight within the band gap regime 
(red line in Fig.~\ref{fig:lw2}(a)).  
The spatial-resolved spectrum reveals that 
these weight mainly come from the boundary of 
the system (Fig.~\ref{fig:lw2}(e)), indicating the existence 
of edge modes within the band gap regime. 

A key feature of the chiral edge mode 
is a unidirectional propagation of spin wave densities.
To confirm this feature, we perform another 
micromagnetic simulation, 
applying a.c. transverse field locally at the boundary of the 
system (red crossed point in Fig.~\ref{fig:mc3}). 
We set an external frequency of  
the a.c. field within the band gap regime; $\omega=6.54$GHz. 
Fig.~\ref{fig:snap} shows several snap shots of the transverse 
magnetization ($m_x(X,Y,t)$) taken after the a.c. field 
is applied from $t=0$. The snap shots clearly demonstrate  
a unidirectional propagation of spin densities in the counterclockwise 
direction. The direction of the propagation is consistent 
with the sign of the group velocity of the chiral edge modes proposed 
in the preceding sections. From the snap shots, the group 
velocity can be estimated to be one unit cell ($a$; linear dimension 
of the unit cell) per 10 ns, which is on the same order 
of the band gap  divided by $2\pi/a$ (the gap $\sim$ 0.2GHz).
The phase velocity of the edge mode is 10 times faster 
than the group velocity and its sign sometimes becomes opposite 
to that of the group velocity (Fig.~\ref{fig:snap}(d)). This  
observation is also consistent with the chiral spin edge mode
proposed in the ring model; the chiral dispersion goes across 
the first Brillouin zone once (Fig.~\ref{fig:1.5-2}(a,b)), 
so that the sign of the phase velocity can be either same or 
opposite to the group velocity.  

\section{Summary and Discussion}

\subsection{summary of our findings}
In this paper, we theoretically explored a realization of topological 
chiral edge mode for magnetostatic spin wave 
in patterned magnetic thin films,   
where magnetic clusters (either rings 
or disks) form a two-dimensional 
square lattice. Without external 
magnetic field, the ground-state spin 
configuration takes a form of circular vortices within each 
ring or disk, respecting the 
square-lattice translational symmetry. Due to the 
magnetic shape anisotropy, spin-wave excitations 
are fully gapped at the zero field. 
When an out-of-plane magnetic 
field is increased up to a saturation field, forward  
spin-wave modes within each ring or disk become 
significantly softened. With the four-fold 
rotational symmetry of the square lattice, 
these modes can be regarded as 
either $s$-wave, $p_x\pm ip_y$-wave or 
$d_{x^2-y^2}$-wave-like `atomic orbitals'. 
When inter-cluster transfer integrals among 
these orbital wave functions are larger than 
frequency spacings among 
their atomic orbital levels, the band-inversion 
between the parity-even 
atomic orbital level ($s$-wave or $d$-wave) and 
parity-odd orbital level ($p_{\pm}$-wave) leads 
to a chiral volume-mode bands with finite Chern integers.  
This results in a chiral (counterclockwise) edge mode 
within a band gap for the volume-mode bands. 

When the system is fully polarized by the out-of-plane field, 
a strong four-fold rotational anisotropy of the 
demagnetization coefficient leads to another effective 
two-bands model. The model is composed of soft magnons  
localized on the nearest neighbor $x$-link 
and that on the $y$-link. Since atomic orbital 
levels for these two are 
same due to the square-lattice symmetry, 
transfer integrals between neighboring soft magnons 
immediately lead to a band inversion mechanism. %As a result, 
The two-orbital model has massive Dirac cone 
like spectra at two inequivalent $X$-points, 
inside which a chiral (counterclockwise) edge mode appears. 
The massive Dirac spectra and the 
edge mode persist for a wide range 
above the saturation field. This feature is 
also justified by micromagnetic simulations. 
  
\subsection{applications to other systems}

\begin{figure}[ht]
   \includegraphics[width=75mm]{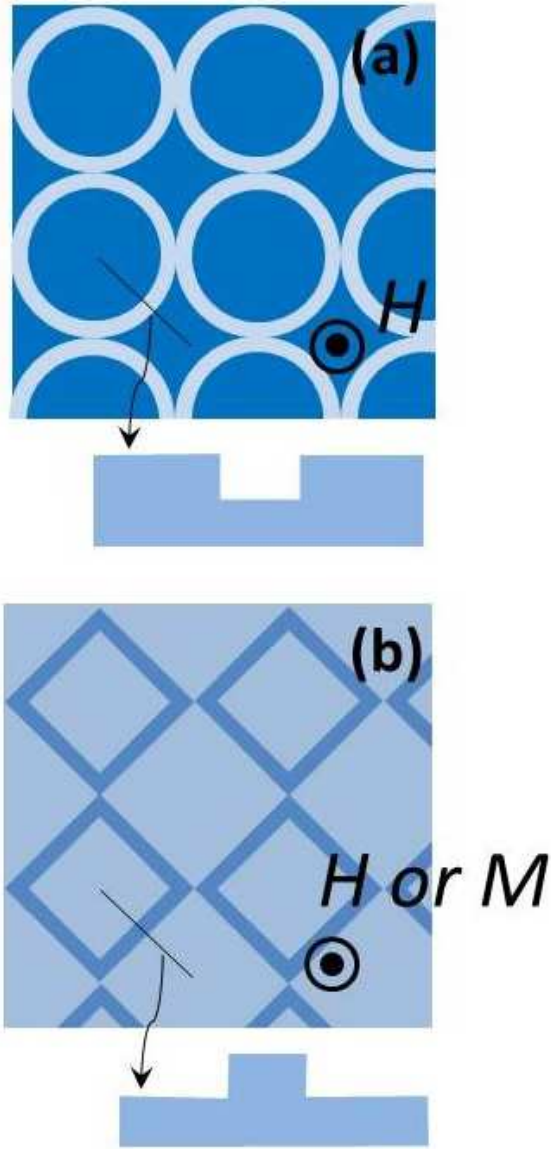}
\caption{ (Color online) Patterned magnetic films with periodically 
aligned gutters (a) or cambers (b). Without 
magnetic crystalline anisotropy (MCA), spins 
at the thinner regions feel stronger easy-plane anisotropy 
than those at that thicker regions.} 
\label{fig:recession}
\end{figure}

In reality, the square-lattice models studied in this paper 
could be placed on some magnetic substrates. 
Also, it is experimentally much easier to engrave only 
a surface of a plane thin film with some periodic  
structuring.~\cite{Adeyeye,Gulyaev} 
The arguments employed in this paper can 
be also applicable to such systems. 
For example, consider that a surface of a magnetic film 
has a number of gutters/cambers forming a square 
lattice, Fig.~\ref{fig:recession}(a)/(b) respectively. 
Due to the magnetostatic energy, moments in thinner film 
regions have stronger easy-plane anisotropy than 
those in thicker film regions. 
Therefore, on applying and increasing an out-of-plane field, 
the moments in thinner regions are expected to 
become fully polarized along the field 
at the highest saturation field, while those in 
the thicker regions do so at the lowest saturation 
field. This means that, in a system shown  
in Fig.~\ref{fig:recession}(a), magnons at the 
gutter region becomes softened  
around the highest saturation field, forming 
atomic orbital wave functions. 
In the other system shown in 
Fig.~\ref{fig:recession}(b), soft modes  
near the lowest saturation field are from the camber region.  
In the presence of the four-fold-rotational symmetry,  
these orbital wave functions play the role of  
either parity-even ($d_{x^2-y^2}$ or 
$s$-waves) orbitals and parity-odd ($p_{\pm}$-waves) 
orbitals, or the in-phase orbitals localized on the 
nearest neighbor $\mu$-link ($\mu=x,y$).  
Thus, provided that neighboring gutters/cambers are 
proximate to each other, the band inversion 
mechanisms described in this paper are expected to be valid, 
leading to a band gap of soft volume-mode 
bands with a chiral (counterclockwise) edge mode.   

The argument is also applicable to thin film ferromagnetic 
materials with 
perpendicular magnetic anisotropy (PMA), where  
relative strength between magnetic shape anisotropy and 
magnetic crystalline anisotropy (MCA) is controlled by the 
film thickness.~\cite{Hubert} In an ultrathin film limit (several atomic 
monolayer), the MCA with easy-axis (out-of-plane) 
anisotropy dominates over the magnetostatic 
energy with easy-plane anisotropy, so that magnetic 
moments are polarized vertically to the plane. 
It has been experimentally 
known that increasing 
film thickness leads to spin-reorientation transition 
from out-of-plane magnetization to in-plane magnetization, 
which indicates that magnetic shape anisotropy overcomes 
the MCA in thicker region.~\cite{Qiu} 
Around the critical thickness, gapped 
spin wave modes are expected to become 
significantly softened.

Regarding the film thickness as alternative to 
the magnetic field, one could also realize topological 
chiral spin-wave edge modes {\it without any external 
magnetic field}. 
For example, consider that a surface of a thin-film 
PMA material is engraved with cambers 
with a lattice periodicity as in Fig.~\ref{fig:recession}(b). 
Suppose that a film-thickness of the camber region 
is chosen near the critical thickness of the material, so that  
magnons around camber regions are sufficiently softened, 
forming orbital wave functions such as $s$, $p_{\pm}$, $d$-waves  
or in-phase orbitals. 
When neighboring cambers are put in close contact with 
one another as in Fig.~\ref{fig:recession}(b), 
exchange processes due to magnetic dipole 
interaction give rise to considerable transfer integrals 
among these orbital wave functions. Although their 
atomic orbital levels within each camber could be also modified 
by the MCA energy, 
we can still expect that larger transfer integrals induce 
the similar type of the band inversion as discussed in 
this paper.

\subsection{a possible experimental method for detecting the 
chiral spin-wave edge mode}
The proposed chiral edge modes 
can be experimentally detected in terms of 
two coils put along the 
boundary of the 2-$d$ magnetic superlattice; 
one is for an input and the other for an 
output (see Fig.~\ref{fig:1.2-1}(c)). 
They are spatially separated 
by tens of the unit cell (for example, 30 unit cells; 0.3mm 
for a unit cell of 10$\mu$m size).  An a.c. electric 
current in the input coil induces an a.c. magnetic field 
near the coil, exciting spin waves (electric 
input). When a frequency of the a.c. current is 
chosen within the band gap regime, the chiral edge mode will 
be selectively excited. The excited spin wave propagates   
along the chiral edge mode and reaches the output 
coil after a certain time delay (e.g. 0.3$\mu$s according 
to the simulation in sec.~IV). When the spin wave reaches 
around the output coil, an a.c. electric current with the 
same frequency will be induced in the output coil 
(electric detection). When the two coils are 
exchanged, the spin wave never reaches the output 
coil, unless it could propagate all the way around the 
boundary without being dissipated. 

When the thickness of the 2-$d$ magnetic 
superlattice is much larger than short-range 
exchange interaction length, the input a.c. current 
can excite not only the proposed topological chiral 
edge modes but also the conventional chiral surface mode; 
Damon-Eshbach (DE) surface mode. In such a case, the 
output a.c. current comprises of two contributions; 
one from the topological chiral edge mode and the 
other from the DE mode. In general, these two modes 
have a number of quantitatively different features. First of 
all, these modes have quite different group velocities; 
the group velocity of the topological edge mode linearly 
depends on the superlattice unit cell size, while that of 
the DE mode doesn't depend on the unit cell size. 
The topological edge mode has a resonance 
frequency within a band gap regimes for volume mode 
bands, which is determined by the magnetic 
superlattice. On the one hand, a resonance frequency 
regime of the DE mode is determined only by the 
out-of-plane magnetization $M$ and field $H$,  
$H<\omega<\sqrt{H(H+4\pi M)}$. When the external 
frequency is changed within the band gap regime, 
the phase velocity of the topological 
mode often changes its sign (see Fig.~\ref{fig:1.5-2}(a,b)), 
while that of DE mode doesn't. In actual experiments, we 
can exploit these distinct features, so as to distinguish 
the contribution of the topological edge mode from that 
of the conventional DE surface mode. 
For example, we can easily differentiate these two 
contributions {\it in time}, by changing the distance between 
the input and output coils. We can further reduce  
one or the other, by changing an external frequency 
of the input a.c. current. Also, by changing the external 
frequency within the band gap regime, we can see the 
phase velocity of the topological mode change its sign.   

\begin{acknowledgements}
The author acknowledges S. Murakami, E. Saitoh, 
G. Tatara, Y. Otani, Y. Fukuma, S. Kasai, Y. Suzuki, 
S. Miwa, Z. Q. Qiu, J. Shi for discussions and informations. 
This work was partly supported by Grant-in-Aids 
from the Ministry of Education, Culture, Sports, 
Science and Technology of Japan 
(Grants No. 21000004, No. 24740225).  
\end{acknowledgements}

\appendix 

\section{Holstein-Primakoff approximation and 
topological Chern integer for magnetostatic spin waves}
In this paper, we considered that magnetic clusters, either thin 
rings or circular disks, form a 2-$d$ periodic lattice; magnetic 
superlattice. To study their magnetostatics and dynamics, 
we used discrete spin models; each cluster is discretized into 
many spins with small volume element, where the spins are 
coupled with one another only via magnetic dipole-dipole interaction. 
We first minimize the magnetostatic energy of the discrete 
spin models, 
\begin{align}
E &\equiv -\frac12 \big(\Delta V\big)^2 
\sum^{i\ne j}_{i,j} \sum_{a,b=x,y,z} 
M_a({\bm r}_i) f_{ab}({\bm r}_i-{\bm r}_j) 
M_b({\bm r}_j)  \nn \\
&\hspace{0.5cm} 
- H\Delta V \sum_i M_z({\bm r}_i), \label{magsta}
\end{align}   
to determine a classical spin configuration ${\bm M}_{0}({\bm r})$.
${\bm r}_i$ specifies a spatial location of a ferromagnetic 
spin with fixed size of moment $|{\bm M}({\bm r}_i)|=M_s$.  
$f_{ab}({\bm r}_i-{\bm r}_j)$ is the magnetic 
dipole-dipole interaction between spin at ${\bm r}_i$ and 
spin at ${\bm r}_j$; 
\begin{eqnarray}
f_{ab}({\bm r}) \equiv -\frac{1}{4\pi} \Big(\frac{\delta_{a,b}}{|{\bm r}|^3}
-\frac{3r_a r_b}{|{\bm r}|^5}\Big). \nn 
\end{eqnarray}
$\Delta V$ denotes a volume element for each spin, whose 
linear dimension is of the same order of short-ranged  
exchange length $l_{\rm ex}$; For YIG and Iron, 
$l_{\rm ex}=18.4$ nm and $2.9$ nm respectively. 

Without the field, the energetically stable  
spin configuration is an array of circular magnetic 
vortices,~\cite{Hubert,Cowburn,Shinjo} respecting the 
periodicity of the square lattice.  Under 
the out-of-plane field, the configurations 
acquire finite out-of-plane moments, which will be 
fully polarized above a saturation field.   
To obtain spin-wave modes, we linearize 
the corresponding Landau-Lifshitz equation in 
favor of fluctuation fields around the classical spin 
configuration. 

In the discrete spin models, the Landau-Lifshitz 
equation take a form of, 
\begin{align}
\partial_t M_a({\bm r}_i) 
&= \epsilon_{abc} \Big[ - H \delta_{b,z} \nn \\ 
& \hspace{1cm} 
- \Delta V \sum_{j\ne i} f_{bd} ({\bm r}_i-{\bm r}_j) 
M_{d}({\bm r}_j) \Big] M_{c}({\bm r}_i). \nn 
\end{align}
Note that the right hand side suggests that 
the saturation field and characteristic spin-wave resonance 
frequency are scaled as $M_s \Delta V/l^3$. Here $l$ 
denotes a distance between the nearest 
neighbor spins in the discrete spin models and 
$1/l^3$ comes from the dipole-dipole interaction between them. 
The small volume element for each spin should be spatially 
isotropic, such that the discrete spin models can 
approximately describe the Maxwell 
equation for magnetic continuum 
media. This requires $\Delta V\simeq l^3$. As a result, 
characteristic spin-wave resonance frequencies 
and saturation field are scaled only 
by the saturation magnetization of a 
constituent material. 
      
The equation of motion is linearized with respect to 
a small transverse field ${\bm m}_{\perp}({\bm r})$ with 
${\bm m}_{\perp}({\bm r}) \equiv {\bm M}({\bm r}) - {\bm M}_{0}({\bm r})$ 
and ${\bm m}_{\perp}({\bm r}) \perp {\bm M}_{0}({\bm r})$. 
With a local spin frame 
in which the classical configuration ${\bm M}_0({\bm r})$ 
becomes fully polarized along the $z$-direction, i.e.  
${\bm R}({\bm r}){\bm M}_{0}({\bm r})=M_s {\bm e}_z$ 
and ${\bm R}({\bm r}){\bm m}_{\perp}({\bm r}) 
= {\bm m}({\bm r})$, the two transverse moments  
in the rotated frame 
${\bm m}({\bm r})=(m_x({\bm r}),m_{y}({\bm r}))$ 
comprise creation/annihilation 
operator for spin wave (magnon);
\begin{align} 
m_{\mp}({\bm r}) \equiv m_x({\bm r}) \pm i m_y({\bm r}). \nn    
\end{align}
With this magnon field, the linearized equation reduces 
to a generalized Hermitian eigenvalue problem, 
\begin{eqnarray}
i\partial_t  
\left(\begin{array}{c}
m_{-}({\bm r}_i) \\
m_{+}({\bm r}_i) \\
\end{array}\right) 
= \sum_{j} {\bm \sigma}_3 \!\ 
\big({\bm H}\big)_{{\bm r}_i,{\bm r}_j} 
\left(\begin{array}{c}
m_{-}({\bm r}_j) \\
m_{+}({\bm r}_j) \\
\end{array}\right), \label{hami0}
\end{eqnarray}
${\bm \sigma}_3$ is a diagonal matrix 
which takes $+1$ in the particle space 
($m_{+}$) and $-1$ in the hole space 
($m_{-}$), reflecting the fact that the 
magnon obeys the bose statistics. 
In this particle-hole space, 
the Hermite matrix is given by   
the following $2$ by $2$ matrix, 
\begin{align}
\big({\bm H}\big)_{{\bm r}_i,{\bm r}_j} 
& \equiv - M_s \alpha({\bm r}_i) \delta_{{\bm r}_i,{\bm r}_j} 
\left(\begin{array}{cc}
1 & \\
& 1 \\
\end{array}\right) \nn \\
&\hspace{-1.2cm} 
- M_s \Delta V (1-\delta_{{\bm r}_i,{\bm r}_j}) 
\left(\begin{array}{cc} 
f_{++}({\bm r}_i,{\bm r}_j) & f_{+-}({\bm r}_i,{\bm r}_j) \\
f_{-+}({\bm r}_i,{\bm r}_j) & f_{--}({\bm r}_i,{\bm r}_j) \\
\end{array}\right). \label{hami} 
\end{align}
$\alpha({\bm r}_i)$ denotes the 
demagnetization coefficient including the static 
out-of-plane field component;
\begin{eqnarray}
\alpha({\bm r}_i) {\bm M}_{0}({\bm r}_i) 
= - \Delta V \sum_{j\ne i} {\bm f}({\bm r}_i
-{\bm r}_j) {\bm M}_0({\bm r}_j) - H {\bm e}_z, \nn
\end{eqnarray} 
where the equality holds true provided that 
the classical spin configuration gives a local 
minimum of the magnetostatic energy, eq.~(\ref{magsta}). 
$f_{\mu\nu}({\bm r}_i,{\bm r}_j)$ ($\mu=\pm$) 
in eq.~(\ref{hami}) represents `exchange' process between 
${\bm r}_i$ and ${\bm r}_j$, which gives rise to propagation 
of magnon excitation under a background of the classical 
spin configuration. The $2$ by $2$ matrix is defined as   
\begin{align}
&\left(\begin{array}{cc}
f_{++}({\bm r},{\bm r}') & f_{+-}({\bm r},{\bm r}') \\
f_{-+}({\bm r},{\bm r}') & f_{--}({\bm r},{\bm r}') \\
\end{array}\right) \nn \\
& \ = \frac12 \left(\begin{array}{cc}
1 & i \\
1 & -i \\
\end{array}\right) \left(\begin{array}{cc}
f_{xx}({\bm r},{\bm r}') & f_{xy}({\bm r},{\bm r}') \\
f_{yx}({\bm r},{\bm r}') & f_{yy}({\bm r},{\bm r}') \\
\end{array}\right) 
\left(\begin{array}{cc}
1 & 1 \\
-i & i \\
\end{array}\right). \label{++}
\end{align} 
$f_{\alpha\beta}({\bm r},{\bm r}')$ ($\alpha,\beta=x,y,z$) 
in the right hand side denotes the dipolar interaction in 
the rotated frame,
\begin{eqnarray}
{\bm f}({\bm r},{\bm r}') 
\equiv {\bm R}({\bm r}){\bm f}({\bm r}-{\bm r}') 
{\bm R}^{t}({\bm r}'). \label{rotated-frame}
\end{eqnarray}

To begin with, consider spin-wave exctitations in a 
circular ring. We treat the ring as a 
one-dimensional chain of many spins which are 
equally spaced from respective neighborings and 
spins along the ring is parameterized by an angle i.e. 
${\bm r}_j=r(\cos\theta_j,\sin\theta_j,0)$ with 
$\theta_j=\frac{2\pi j}{M}$ $(j=1,2,\cdots,M)$. 
`$r$' denotes the radius of the ring. 
The classical spin configuration minimizing 
the magnetostatic energy eq.~(\ref{magsta}) respects 
the circular symmetry,
\begin{align}
& {\bm M}_{0}({\bm r}_j) = 
M_s (-\sin\varphi\sin\theta_j,\sin\varphi\cos\theta_j,\cos\varphi). 
\nn
\end{align} 
$\varphi$ denotes a relative angle between each spin and 
the external magnetic field, which is independent 
from $j$ due to the circular symmetry. To introduce a magnon  
and its Hamiltonian in a ring, we take a following 
local spin frame in eqs.~(\ref{hami}-\ref{rotated-frame}), 
\begin{align}
{\bm R}({\bm r}_j) = \left(\begin{array}{ccc}
1 & & \\
& \cos\varphi & -\sin\varphi \\
& \sin\varphi & \cos\varphi \\
\end{array}\right) 
\left(\begin{array}{ccc}
\cos\theta_j & \sin\theta_j & \\
-\sin\theta_j & \cos\theta_j & \\
& & 1 \\
\end{array}\right). \nn
\end{align} 
Under this gauge, the right hand side of 
eq.~(\ref{rotated-frame}) depends only on a relative 
angle between two magnetic elements along the 
ring;
\begin{widetext}
\begin{align}
{\bm f}({\bm r}_i,{\bm r}_j) 
&={\bm f}(\theta_i-\theta_j) 
= - \frac{1}{32\pi a^3 |\sin\frac{\theta_i-\theta_j}{2}|^3} 
\Bigg\{\left(\begin{array}{ccc}
c_{\theta_i-\theta_j} & s_{\theta_i-\theta_j} c_{\varphi}  & 
s_{\theta_i-\theta_j} s_{\varphi} \\
- s_{\theta_i-\theta_j} c_{\varphi} & c_{\theta_i-\theta_j} 
c^2_{\varphi} + s^2_{\varphi} & (c_{\theta_i-\theta_j} -1)c_{\varphi}s_{\varphi} \\
- s_{\theta_i-\theta_j}s_{\varphi} & (c_{\theta_i-\theta_j}-1) c_{\varphi}s_{\varphi} 
& s^2_{\varphi} c_{\theta_i-\theta_j} + c^2_{\varphi} \\
\end{array}\right)  \nn \\
& \hspace{3cm} 
- \frac{3}{2} 
\left(\begin{array}{ccc}
-(1-c_{\theta_i-\theta_j}) & 
s_{\theta_i-\theta_j} c_{\varphi}  &  s_{\theta_i-\theta_j} s_{\varphi} \\
- s_{\theta_i-\theta_j} c_{\varphi} & (c_{\theta_i-\theta_j} -1) 
c^2_{\varphi}  & (c_{\theta_i-\theta_j}-1) c_{\varphi}s_{\varphi} \\
- s_{\theta_i-\theta_j}s_{\varphi} & c_{\varphi}s_{\varphi} 
(c_{\theta_i-\theta_j}-1) & 
s^2_{\varphi} (c_{\theta_i-\theta_j}-1) \\
\end{array}\right) \Bigg\}
\end{align}
\end{widetext}
with $c_{\theta_i-\theta_j}\equiv \cos(\theta_i-\theta_j)$, 
$s_{\theta_i-\theta_j} \equiv \sin(\theta_i-\theta_j)$, 
$c_{\varphi} \equiv \cos\varphi$, and $s_{\varphi} 
\equiv \sin\varphi$. The demagnetization coefficient 
in an isolated ring also respects the circular symmetry; 
$\alpha({\bm r}_j)=\alpha$. Thus, the magnon 
Hamiltonian for a circular ring depends only on the 
relative angle;
\begin{eqnarray}
i\partial_t  
\left(\begin{array}{c}
m_{-}(\theta_i) \\
m_{+}(\theta_i) \\
\end{array}\right) 
= \sum^{M}_{j=1} {\bm \sigma}_3 \!\ 
\big({\bm H}\big)_{\theta_i-\theta_j} 
\left(\begin{array}{c}
m_{-}(\theta_j) \\
m_{+}(\theta_j) \\
\end{array}\right), \label{hami0}
\end{eqnarray} 
Correspondingly, the spin-wave excitations in a 
ring are characterized by the angular momentum 
variable $q_J=\frac{2\pi n_J}{M}$ $(n_J=-\frac{M}{2},-\frac{M}{2}+1
,\cdots,\frac{M}{2})$ 
associated with the circular symmetry;
\begin{eqnarray}
\left(\begin{array}{c}
m_{-}(\theta_i) \\
m_{+}(\theta_i) \\
\end{array}\right)  = 
\sum_{n_J} e^{in_J \theta_i} \left(\begin{array}{c}
m_{-}(n_J) \\
m_{+}(-n_J) \\
\end{array}\right).   
\end{eqnarray}
The linearized equation is given by   
\begin{eqnarray}
i\partial_t  
\left(\begin{array}{c}
m_{-}(n_J) \\
m_{+}(-n_J) \\
\end{array}\right) 
= {\bm \sigma}_3 \!\ \big({\bm H}\big)_{n_J} 
\left(\begin{array}{c}
m_{-}(n_J) \\
m_{+}(-n_J) \\
\end{array}\right),
\end{eqnarray}
with 
\begin{eqnarray}
({\bm H})_{n_J} \equiv \sum_{j} e^{in_J \theta_j} 
({\bm H})_{\theta_j}. \nn
\end{eqnarray} 
The 2 by 2 Hermite matrix  $({\bm H})_{n_J}$ is diagonalized 
for each angular momentum in terms of canonical transformation 
(2 by 2 paraunitary matrix);
\begin{eqnarray}
&({\bm H})_{n_J} 
{\bm t}_{n_J}  = {\bm \sigma}_3 \!\ 
{\bm t}_{n_J} E_{n_J}.  
\end{eqnarray}  
with a proper normalization 
${\bm t}^{\dagger}_{n_J}{\bm \sigma}_3{\bm t}_{n_J}=1$. 
Positive definite $E_{n_J}$ stands for a resonance frequency 
for the spin-wave excitations in a circular ring. Respective 
spin-wave mode is represented by the two-component vector 
in the particle-hole space ${\bm t}_{n_J}$; the linearized 
equation of motion eq.~(\ref{hami0}) is satisfied by  
\begin{align}
\psi_{q_J}(\theta_j) = {\bm t}_{n_J} \!\ e^{in_J \theta_j-iE_{n_J} t}  
\label{psi-def}
\end{align}
with $q_J\equiv \frac{2\pi n_J}{M}$. 
Eq.~(\ref{psi-def}) with $n_J=-\frac{M}{2},\cdots,\frac{M}{2}-1,\frac{M}{2}$ 
comprise `atomic orbital' wavefunctions within a circular ring, which 
are classified by the total angular momentum $q_J$;
\begin{eqnarray}
\psi_{q_J}(\theta_j+\theta_m) = e^{iq_J m} \psi_{q_J}(\theta_j). \nn
\end{eqnarray}   
These wavefunctions gives us bases for   
tight-binding descriptions of spin-wave excitations in 
the magnetic superlattice.

To obtain spin-wave dispersion relations 
for volume modes and edge modes in the magnetic 
superlattices, we 
diagonalize eq.~(\ref{hami}) with a periodic boundary 
condition along the $x$-direction and an open boundary 
condition along the $y$-direction. A system  
typically contains $9$-$18$ square-lattice unit cell along 
the $y$-direction. We minimize the magnetostatic energy,  
respecting the periodicity of the square lattice, 
${\bm M}_0({\bm r}+{\bm e}_x)={\bm M}_0({\bm r})$. So do 
$\alpha({\bm r})$, ${\bm R}({\bm r})$ and 
$({\bm H})_{{\bm r}_i,{\bm r}_j}$; 
$({\bm H})_{{\bm r}_i+{\bm e}_{x},{\bm r}_j}=
({\bm H})_{{\bm r}_i,{\bm r}_j-{\bm e}_{x}}$. Correspondingly,  
we diagonalize the following fourier-transformed Hamiltonian,
\begin{align}
({\bm H}_{k})_{{\bm r}_i,{\bm r}_j} 
&= - M_s \alpha({\bm r}_i) \delta_{{\bm r}_i,{\bm r}_j} \nn \\ 
&\ \  \ - M_s \Delta V \left(\begin{array}{cc}
f_{k,++}({\bm r}_i,{\bm r}_j) & f_{k,+-}({\bm r}_i,{\bm r}_j) \\
f_{k,-+}({\bm r}_i,{\bm r}_j) & f_{k,--}({\bm r}_i,{\bm r}_j) \\
\end{array}\right), \label{hermitian} 
\end{align}  
with 
\begin{align}
f_{k,\sigma\sigma'}({\bm r},{\bm r}')
&= e^{-i k({\bm r}-{\bm r}')_x} \times \nn \\
&\ \ \ \sum_{\bm b} (1-\delta_{{\bm r},{\bm r}'-{\bm b}})
f_{\sigma\sigma'}({\bm r},{\bm r}'-{\bm b}) e^{-i k{\bm b}_x}. \nn
\end{align}
with $\sigma,\sigma'=\pm$. The summation over 
the lattice translational vector ${\bm b}$ is 
taken only along the $x$-direction and is over 
a finite range ${\bm b}\equiv n{\bm e}_x$ 
with $-10\le n\le 10$. 
Provided that the classical spin configuration 
${\bm M}_0({\bm r}_i)$ gives a local minimum 
for the magnetostatic energy, 
the linearized Hamiltonian is paraunitarily 
equivalent to a positive definite 
diagonal matrix ${\bm E}_{k}$: ${\bm T}^{\dagger}_{k} 
{\bm H}_{k} {\bm T}_{k} = 
{\bm E}_{k}$ with ${\bm T}^{\dagger}_{k}{\bm \sigma}_3 {\bm T}_{k} 
= {\bm \sigma}_3$. Each diagonal 
element in ${\bm E}_k$ and corresponding column vector 
in ${\bm T}_{k}$ gives a resonance frequency 
and wave function for a volume mode and edge mode as a 
function of  the wave vector $k$ along the $x$-direction. 
With the normalization condition of ${\bm T}_k$ in mind, 
an amplitude of the wave function for the $n$-th 
eigen mode at ${\bm r}_i$ is defined as 
$\sum_{\sigma=\pm} \sigma|({\bm T}_k)_{({\bm r}_j,\sigma|n)}|^2$.  
We regard the mode as an edge mode, when more 
than $70\%$ of the amplitude is localized along 
the boundaries of the system (see also the captions of 
Figs.~\ref{fig:1.2-1}, \ref{fig:cir-band}). Otherwise, we observed 
that wave functions are usually extended over the system, 
and thus can be regarded as volume modes.       

Dispersion relations for the volume modes are also 
obtained from calculations with periodic boundary conditions 
imposed on both $x$ and $y$-direction. The classical ground-state
spin configuration respects the 
periodicity of the square lattice, 
${\bm M}_0({\bm r}+{\bm e}_x)={\bm M}_0({\bm r}+{\bm e}_y) 
= {\bm M}_0({\bm r})$. We 
diagonalize eq.~(\ref{hermitian}) with $k$ being 
replaced by ${\bm k}=(k_x,k_y)$, in terms of a paraunitary 
transformation ${\bm T}_{\bm k}$;
\begin{align}
f_{{\bm k},\sigma\sigma'}({\bm r},{\bm r}')
&= e^{-i {\bm k}({\bm r}-{\bm r}')} \times \nn \\
&\ \ \ \sum_{\bm b} (1-\delta_{{\bm r},{\bm r}'-{\bm b}})
f_{\sigma\sigma'}({\bm r},{\bm r}'-{\bm b}) e^{-i {\bm k}{\bm b}}, \nn
\end{align} 
with ${\bm b}=n{\bm e}_x+m{\bm e}_y$ and $-10 \le n,m \le 10$.      
The topological Chern integer for the $j$-th volume mode band  
is defined by the $j$-the column vector of the paraunitary 
matrix ${\bm T}_{\bm k}$ as 
\begin{eqnarray}
c_j \equiv \frac{i \epsilon_{\mu\nu}}{2\pi} \int_{\rm BZ} d^2{\bm k} 
\!\ {\rm  Tr} \Big[{\bm \Gamma}_j {\bm \sigma}_3 
\big(\partial_{k_\mu}{\bm T}^{\dagger}_{\bm k}\big) 
{\bm \sigma}_3 
\big(\partial_{k_\nu}{\bm T}_{\bm k}\big) \Big]. \nn 
\end{eqnarray} 
Here ${\bm \Gamma}_j$ takes $+1$ in the $(j,j)$ component 
while $0$ otherwise. $c_j$ takes an integer 
and describes a topological structure of a wave function 
for the $j$-th volume mode band in the two-dimensional 
Brillouin zone (BZ).~\cite{SO1,TKKN,Avron}   

\section{two-orbital model valid above the saturation field}
\begin{figure}[ht]
   \includegraphics[width=70mm]{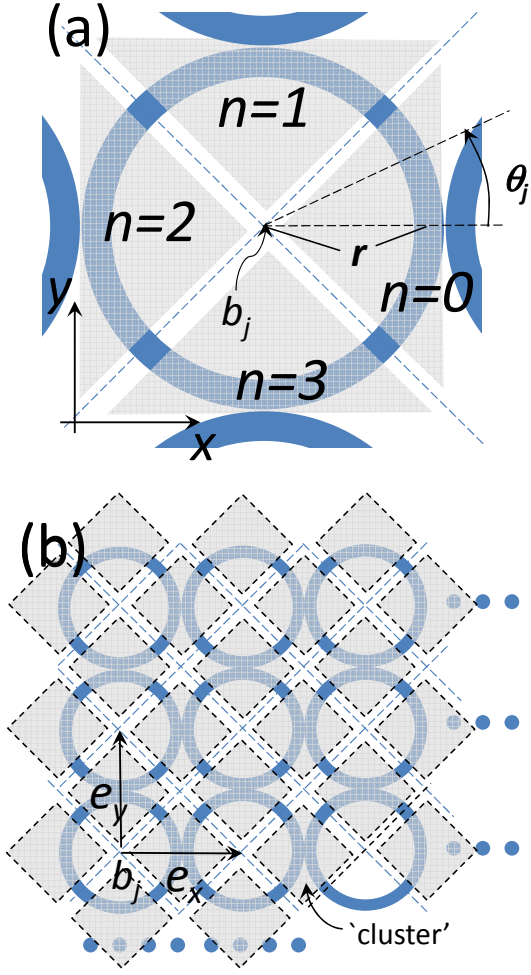}
\caption{ (Color online) (a) Ring is decomposed into 
four quadrants (grey shadow regions), which are ranged as 
$-\frac{\pi}{4}\le \theta_j \le \frac{\pi}{4}$ ($n=0$), 
$\frac{\pi}{4}\le \theta_j \le \frac{3\pi}{4}$ ($n=1$), 
$\frac{3\pi}{4}\le \theta_j \le \frac{5\pi}{4}$ ($n=2$), 
$\frac{5\pi}{4}\le \theta_j \le \frac{7\pi}{4}$ ($n=3$)  
respectively with ${\bm r}_j={\bm b}_j+r(\cos\theta_j,\sin\theta_j)$. 
Here ${\bm b}_j$ denotes a center of the ring and 
$r$ is a radius of the ring. (b) One quadrant in a ring 
and its closest quadrant in the nearest neighbor ring 
are combined together, to 
form a cluster (grey shadow region encompassed 
by a black dotted line). The cluster thus defined is centered 
at ${\bm b}_j+\frac{{\bm e}_x}{2}$ (a mid-point of the 
nearest neighbor $x$-link) or 
${\bm b}_j+\frac{{\bm e}_y}{2}$ (a mid-point of the $y$-link), 
where ${\bm e}_x$ and ${\bm e}_y$ are the 
basic translational vectors.} 
\label{fig:cluster}
\end{figure}

When the classical spin configuration is fully polarized 
along the out-of-plane field, demagnetization field at the four 
corner of a ring, ${\bm r}_j={\bm b}_j+r(\cos\theta_j,\sin\theta_j)$ 
with $\theta_j=0,\frac{\pi}{2},\pi,\frac{3\pi}{2}$, 
is much stronger than those in the others. Here ${\bm b}_j$ 
denote a coordinate of a center of the ring at which a spin 
at ${\bm r}_j$ is included. $r$ is   
a radius of the ring (Fig.~\ref{fig:cluster}). As a result, 
soft volume mode bands are mainly composed of   
spins localized at $\theta_j=0,\frac{\pi}{2},\pi,\frac{3\pi}{2}$. 
In such a case, exchange process between nearest neighbor 
rings becomes even larger than that within a same ring.  
We thus take into account the former exchange 
process first, to introduce {\it atomic orbital wave functions   
defined on a link connecting two nearest neighboring rings}.

Specifically, we first decompose every ring,  
${\bm r}_j={\bm b}_j+r(\cos\theta_j,\sin\theta_j)$,  
into four quadrants, which are ranged as $-\frac{\pi}{4}\le \theta_j 
\le \frac{\pi}{4}$,  $\frac{\pi}{4} \le \theta_j \le \frac{3\pi}{4}$, 
$\frac{3\pi}{4} \le \theta_j \le \frac{5\pi}{4}$, 
$\frac{5\pi}{4} \le \theta_j \le \frac{7\pi}{4}$ respectively 
(Fig.~\ref{fig:cluster}(a)). We 
then combine one quadrant in a ring 
($-\frac{\pi}{4}\le \theta_j 
\le \frac{\pi}{4}$ with ${\bm b}_j$ or $\frac{\pi}{4}\le \theta_j 
\le \frac{3\pi}{4}$ with ${\bm b}_j$) and its closest quadrant 
of the nearest neighboring ring ($\frac{3\pi}{4}\le \theta_j 
\le \frac{5\pi}{4}$ with ${\bm b}_j+{\bm e}_x$ or 
$\frac{5\pi}{4}\le \theta_j 
\le \frac{7\pi}{4}$ with ${\bm b}_j+{\bm e}_y$ respectively), 
to make a `cluster' (Fig.~\ref{fig:cluster}(b)). 
The cluster thus defined 
is centered at a middle point of the nearest 
neighbor $x$-link or that of the 
$y$-link (${\bm b}_j+\frac{{\bm e}_x}{2}$ or 
${\bm b}_j+\frac{{\bm e}_y}{2}$ respectively). Correspondingly, 
we decompose the BdG Hamiltonian in eq.~(\ref{hami}) 
into two parts, one is diagonal with respect 
to a cluster index and the other is 
off-diagonal with respect to 
the cluster index;  
\begin{align}
\big({\bm H}\big)_{{\bm r}_i,{\bm r}_j} 
= \big({\bm H}_0\big)_{{\bm r}_i,{\bm r}_j} + 
\big({\bm H}_1\big)_{{\bm r}_i,{\bm r}_j}, 
\end{align}
with 
\begin{align}
\big({\bm H}_0\big)_{{\bm r}_i,{\bm r}_j} &= - M_s 
\alpha({\bm r}_i) \delta_{{\bm r}_i,{\bm r}_j} \left(\begin{array}{cc}
1 & \\
& 1 \\
\end{array}\right) - M_s \Delta V \times \nn \\ 
&\hspace{1cm} \delta_{[{\bm r}_i],[{\bm r}_j]} \!\ 
\eta_{{\bm r}_i,{\bm r}_j} \left(\begin{array}{cc}
f_{++}({\bm r}_i,{\bm r}_j) & f_{+-}({\bm r}_i,{\bm r}_j) \\
f_{-+}({\bm r}_i,{\bm r}_j)&  f_{--} ({\bm r}_i,{\bm r}_j) \\
\end{array}\right), \nn \\
\big({\bm H}_1\big)_{{\bm r}_i,{\bm r}_j} &=  - M_s \Delta V 
\!\  \eta_{[{\bm r}_i],[{\bm r}_j]} \left(\begin{array}{cc}
f_{++}({\bm r}_i,{\bm r}_j) & f_{+-}({\bm r}_i,{\bm r}_j) \\
f_{-+}({\bm r}_i,{\bm r}_j)&  f_{--} ({\bm r}_i,{\bm r}_j) \\
\end{array}\right), \nn \\
\end{align}
where $\eta_{{\bm r}_i,{\bm r}_j}\equiv 1- \delta_{{\bm r}_i,{\bm r}_j}$,  
$\eta_{[{\bm r}_i],[{\bm r}_j]}\equiv 1- \delta_{[{\bm r}_i],[{\bm r}_j]}$ 
and $[{\bm r}_i]$ species a cluster in which a spin site ${\bm r}_i$ 
is included.
Now that the spin configuration is fully polarized, 
we take the following frame in eq.~(\ref{rotated-frame}), 
\begin{eqnarray}
{\bm R}({\bm r}_i) \equiv \left(\begin{array}{ccc}
\cos\theta_i & \sin\theta_i & \\
-\sin\theta_i & \cos\theta_i & \\
& & 1 \\
\end{array}\right). \label{rotation2}
\end{eqnarray} 
with ${\bm r}_i\equiv {\bm b}_i+r(\cos\theta_i,\sin\theta_i)$.  
With this rotated spin frame, the $2$ by $2$ transfer 
integrals is given as  
\begin{widetext}
\begin{align}
&\left(\begin{array}{cc}
f_{++}({\bm r}_i,{\bm r}_j) & f_{+-}({\bm r}_i,{\bm r}_j) \\
f_{-+}({\bm r}_i,{\bm r}_j) & f_{--}({\bm r}_i,{\bm r}_j) \\
\end{array}\right) = \frac{1}{4\pi R^3} \!\ \Bigg\{ - 
\left(\begin{array}{cc}
e^{-i(\theta_i-\theta_j)} & \\
& e^{i(\theta_i-\theta_j)} \\
\end{array}\right) 
+ \frac{3}{2} \left(\begin{array}{cc}
e^{-i(\theta_i-\theta_j)}  & e^{-i(\theta_i+\theta_j)+2i\varphi_{ij}} \\
e^{i(\theta_i+\theta_j)-2i\varphi_{ij}} & e^{i(\theta_i-\theta_j)} \\
\end{array}\right) \Bigg\}  \label{rotation}
\end{align} 
\end{widetext}
with ${\bm r}_j \equiv {\bm b}_j + r(\cos\theta_j,\sin\theta_j)$, 
$R\equiv |{\bm r}_i-{\bm r}_j|$ and ${\bm r}_i-{\bm r}_j  
\equiv R (\cos\varphi_{ij},\sin\varphi_{ij})$. 
 In the following, we first diagonalize ${\bm H}_0$ to 
introduce `orbital wave functions' within 
each cluster. In terms of this orbital 
basis, we next include 
${\bm H}_1$ as a inter-cluster transfer integrals. 

To carry out this procedure systematically, 
we further decompose 
the diagonal part into two parts, 
${\bm H}_0 \equiv {\bm H}^{\prime}_0
+ {\bm H}^{\prime\prime}_0$, where  
$({\bm H}^{\prime}_{0})_{{\bm r}_i,{\bm r}_j}$ is no-zero if and only if both 
${\bm r}_i$ and ${\bm r}_j$ are within the same quadrant, 
while $({\bm H}^{\prime\prime}_0)_{{\bm r}_i,{\bm r}_j}$ 
is non-zero if ${\bm r}_i$ is in one quadrant 
and ${\bm r}_j$ is in the other; ${\bm H}^{\prime\prime}_0$ 
plays the part of exchange process between the nearest 
neighbor quadrants.  
Consider first the lowest eigen basis which diagonalizes 
${\bm H}^{\prime}_0$;
\begin{align}
&{\bm H}^{\prime}_{0} |u_{\pm,n,{\bm b}}\rangle  
= {\bm \sigma}_3 |u_{\pm,n,{\bm b}}\rangle (\pm E), 
\label{basis} 
\end{align} 
where $\langle u_{\nu,n,{\bm b}}|{\bm \sigma}_3 
|u_{\mu,n^{\prime},{\bm b}^{\prime}}\rangle  
= \nu \delta_{\nu,\mu} 
\delta_{n,n^{\prime}} \delta_{{\bm b},{\bm b}^{\prime}}$ 
with $\nu,\mu=\pm$. ${\bm b}$ and  
${\bm b}^{\prime}$ denote spatial coordinate of (a center of) 
the ring to which the basis belongs, while the subscripts 
$n,n^{\prime} (=0,1,2,3)$ specify the quadrant to which the 
basis belongs. For example, 
$\langle {\bm r}_j,\tau|u_{\pm,0,{\bm b}}\rangle$ 
is non-zero only when ${\bm b}={\bm b}_j$ and 
$-\frac{\pi}{4}\le \theta_j \le \frac{\pi}{4}$ 
with ${\bm r}_j={\bm b}_j + r(\cos\theta_j,\sin\theta_j)$, 
while $\langle {\bm r}_j,\tau|u_{\pm,1,{\bm b}}\rangle$ 
is non-zero only when ${\bm b}={\bm b}_j$ and 
$\frac{\pi}{4}\le \theta_j \le \frac{3\pi}{4}$ and so on (see 
also fig.~\ref{fig:cluster}(a) for $n=2,3$).   
$|u_{+,n,{\bm b}}\rangle$ and 
$|u_{-,n,{\bm b}}\rangle$ are particle-hole pair to each other, 
\begin{align}
\langle {\bm r}_i,\tau|u_{-,n,{\bm b}}\rangle 
= ({\bm \sigma}_1)_{\tau\tau^{\prime}} 
\langle u_{+,n,{\bm b}}|{\bm r}_i,{\tau}^{\prime}\rangle, \label{p-h} 
\end{align}
with the particle-hole index $\tau=1,2$. Due to the four-fold 
rotational  and square-lattice translational symmetries, 
the lowest eigen frequency in eq.~(\ref{basis}), $E$, 
does not depend on $n$ and ${\bm b}$.

Now that $-\alpha({\bm r}_j)$ in ${\bm H}^{\prime}_0$ has  
deep minima at $\theta_j=0,\frac{\pi}{2},\pi,\frac{3\pi}{2}$ with 
${\bm r}_j={\bm b}_j+r(\cos\theta_j,\sin\theta_j)$, 
the lowest eigen basis is expected to be localized 
around these valley bottoms,  
\begin{align}
&\big\langle {\bm r}_{j}, \tau\big|
u_{+,n,{\bm b}} \big\rangle \simeq \delta_{{\bm b},{\bm b}_j}
\Big(\delta_{n,0}\delta_{\theta_j,0} + 
\delta_{n,1}\delta_{\theta_j,\frac{\pi}{2}} \nn \\
&\hspace{2cm} + \delta_{n,2}\delta_{\theta_j,\pi} + 
\delta_{n,3}\delta_{\theta_j,\frac{3\pi}{2}} \Big) 
\left(\begin{array}{cc}
u \\
v \\
\end{array}\right)_{\tau}. \label{w1}
\end{align}    
$(u,v)$ represents a two-component vector 
in the particle-hole space. 
Near (but above) the saturation field, the  
vector is equally-weighted in the particle-hole space, 
\begin{align}
\left(\begin{array}{c}
u \\
v \\
\end{array}\right) \simeq \left(\begin{array}{c}
i \\
- i \\
\end{array}\right) \ \ \ {\rm for} \ \ H\gtrsim H_c.  \label{eq1}
\end{align}
The relative phase between the particle-component ($u$; $\tau=1$)
and the hole component ($v$; $\tau=2$) was taken $-1$, 
because a condensation of the soft magnon with eq.~(\ref{eq1})    
results in an in-plane component which is {\it tangential} to 
the ring; 
the in-plane component of the classical spin configuration 
at $H<H_c$ takes 
the circular vortex structure within each ring. Note 
also that, in eq.~(\ref{w1}), the relative phase among 
different quadrants was chosen to be $+1$, because of 
the rotated spin frame, eq.~(\ref{rotation2}). In the high 
field limit, the vector is polarized in the particle space, 
\begin{align}
\left(\begin{array}{c}
u \\
v \\
\end{array}\right) \rightarrow 
\left(\begin{array}{c}
1 \\
0 \\
\end{array}\right)   \ \ \ {\rm for} \ \ H\rightarrow \infty. \label{full}
\end{align}

In terms of the lowest eigen basis of ${\bm H}^{\prime}_0$, 
${\bm H}_0$ takes a form; 
\begin{widetext}
\begin{align}
{\bm H}_0 =& \sum_{{\bm b}} \Big\{ 
\left(\begin{array}{cccc} 
\gamma^{\dagger}_{0,{\bm b}} & \gamma^{\dagger}_{2,{\bm b}+{\bm e}_x} 
& \gamma_{0,{\bm b}} & \gamma_{2,{\bm b}+{\bm e}_x} 
\end{array}\right)\left(\begin{array}{cccc} 
E & t & 0 & s \\
t & E & s & 0 \\
0 & s & E & t \\
s & 0 & t & E \\
\end{array}\right) \left(\begin{array}{c}
\gamma_{0,{\bm b}} \\
\gamma_{2,{\bm b}+{\bm e}_x} \\
\gamma^{\dagger}_{0,{\bm b}} \\
\gamma^{\dagger}_{2,{\bm b}+{\bm e}_x} \\
\end{array}\right) \nn \\
&  \hspace{1cm} + \left(\begin{array}{cccc} 
\gamma^{\dagger}_{1,{\bm b}} & \gamma^{\dagger}_{3,{\bm b}+{\bm e}_y} 
& \gamma_{1,{\bm b}} & \gamma_{3,{\bm b}+{\bm e}_y} 
\end{array}\right)\left(\begin{array}{cccc} 
E & t & 0 & s \\
t & E & s & 0 \\
0 & s & E & t \\
s & 0 & t & E \\
\end{array}\right) \left(\begin{array}{c}
\gamma_{1,{\bm b}} \\
\gamma_{3,{\bm b}+{\bm e}_y} \\
\gamma^{\dagger}_{1,{\bm b}} \\
\gamma^{\dagger}_{3,{\bm b}+{\bm e}_y} \\
\end{array}\right)
\Big\} \label{4by4}
\end{align} 
\end{widetext}
where $\gamma^{\dagger}_{n,{\bm b}}$ / $\gamma_{n,{\bm b}}$ 
denotes a creation / annihilation operator 
which excites $|u_{+,n,{\bm b}}\rangle$ / $|u_{-,n,{\bm b}}\rangle$ 
respectively. 
$t$ and $s$ are real-valued and represent hopping 
terms between two nearest neighboring quadrants in 
the particle-particle channel and particle-hole 
channel respectively, 
\begin{align}
t & \equiv \langle u_{+,0,{\bm b}} | 
{\bm H}^{\prime\prime}_{0} | u_{+,2,{\bm b}+{\bm e}_x} \rangle 
    = \langle u_{+,2,{\bm b}+{\bm e}_x} | 
{\bm H}^{\prime\prime}_{0} | u_{+,0,{\bm b}} \rangle \nn \\
 &= \langle u_{-,0,{\bm b}}  | {\bm H}^{\prime\prime}_{0} 
| u_{-,2,{\bm b}+{\bm e}_x} \rangle  
  =  \langle u_{-,2,{\bm b}+{\bm e}_x} |
{\bm H}^{\prime\prime}_{0} | u_{-,0{\bm b}} \rangle , \label{t-ha} 
\end{align}
\begin{align}
s & \equiv \langle u_{+,0,{\bm b}} | 
{\bm H}^{\prime\prime}_{0}  | u_{-,2,{\bm b}+{\bm e}_x} \rangle  
    = \langle u_{+,2,{\bm b}+{\bm e}_x} |  
{\bm H}^{\prime\prime}_{0} | u_{-,0,{\bm b}} \rangle  \nn \\
 &= \langle u_{-,0,{\bm b}} | 
{\bm H}^{\prime\prime}_{0} |   
u_{+,2,{\bm b}+{\bm e}_x} \rangle 
  =  \langle u_{-,2,{\bm b}+{\bm e}_x} |  
{\bm H}^{\prime\prime}_{0} | u_{+,0,{\bm b}} \rangle . \label{s-ha} 
\end{align}   
The equalities in eqs.~(\ref{t-ha},\ref{s-ha}) 
come from the particle-hole symmetry, 
$\pi$-rotational symmetry and a mirror 
symmetry combined with 
the time-reversal. Diagonalization of eq.~(\ref{4by4}) 
introduces orbital wave functions on 
the nearest neighbor $x$-link as,  
\begin{align}
&\left(\begin{array}{c}
\beta^{\dagger}_{-,{\bm b}+\frac{{\bm e}_x}{2}} \\
\beta^{\dagger}_{+,{\bm b}+\frac{{\bm e}_x}{2}} \\
\beta_{-,{\bm b}+\frac{{\bm e}_x}{2}} \\
\beta_{+,{\bm b}+\frac{{\bm e}_x}{2}} \\
\end{array}\right) = \frac{1}{\sqrt{2}} \times \nn \\
&\hspace{0.2cm} 
\left(\begin{array}{cccc}
{\rm ch}_{\frac{\theta}{2}} & {\rm ch}_{\frac{\theta}{2}}  
& {\rm sh}_{\frac{\theta}{2}} &  {\rm sh}_{\frac{\theta}{2}} \\
- {\rm ch}_{\frac{\theta^{\prime}}{2}} & {\rm ch}_{\frac{\theta^{\prime}}{2}} 
& - {\rm sh}_{\frac{\theta^{\prime}}{2}} & {\rm sh}_{\frac{\theta^{\prime}}{2}} \\
{\rm sh}_{\frac{\theta}{2}} & {\rm sh}_{\frac{\theta}{2}}
& {\rm ch}_{\frac{\theta}{2}} &  {\rm ch}_{\frac{\theta}{2}} \\
-{\rm sh}_{\frac{\theta^{\prime}}{2}} & {\rm sh}_{\frac{\theta^{\prime}}{2}} 
& -{\rm ch}_{\frac{\theta^{\prime}}{2}} & 
{\rm ch}_{\frac{\theta^{\prime}}{2}} \\
\end{array}\right)  \left(\begin{array}{c}
\gamma^{\dagger}_{0,{\bm b}} \\
\gamma^{\dagger}_{2,{\bm b}+{\bm e}_x} \\
\gamma_{0,{\bm b}} \\
\gamma_{2,{\bm b}+{\bm e}_x} \\
\end{array}\right) \label{bog} 
\end{align}  
with $({\rm ch}_{\frac{\theta}{2}},{\rm sh}_{\frac{\theta}{2}})
\equiv (\cosh \frac{\theta}{2},\sinh \frac{\theta}{2})$ and  
\begin{align}
&\cosh\theta = \frac{E+t}{\sqrt{(E+t)^2-s^2}}, \!\    
\sinh\theta=\frac{s}{\sqrt{(E+t)^2-s^2}}, \nn \\   
&\cosh\theta^{\prime} = \frac{E-t}{\sqrt{(E-t)^2-s^2}}, \!\
\sinh\theta^{\prime}=\frac{-s}{\sqrt{(E-t)^2-s^2}}. \nn
\end{align}
$\beta_{+,{\bm b}+\frac{{\bm e}_x}{2}}$/$\beta_{-,{\bm b}+\frac{{\bm e}_x}{2}}$ 
is for a `in-phase'/`out-of-phase' 
orbital formed by $\gamma_{0,{\bm b}}$ and 
$\gamma_{2,{\bm b}+{\bm e}_x}$, whose eigen frequency is 
 $\sqrt{(E-t)^2-s^2}$/$\sqrt{(E+t)^2-s^2}$ respectively. 
Under the rotated spin frame, eq.~(\ref{rotation2}), these 
two in fact stand for an    
`in-phase'/`out-of-phase' mode formed 
by a spin at ${\bm r}={\bm b}+(r,0)$ and that at 
${\bm r}={\bm r}+{\bm e}_x-(r,0)$ respectively.  
Similarly, the in-phase/out-of-phase orbitals between 
$\gamma_{1,{\bm b}}$ and $\gamma_{3,{\bm b}+{\bm e}_y}$ 
are introduced on the nearest neighboring 
$y$-link, $\beta_{\pm,{\bm b}+\frac{{\bm e}_y}{2}}$;
\begin{align}
{\bm H}_0 &= 2 \sum_{{\bm b}} \sum_{\mu=x,y} 
\Big\{  \sqrt{(E-t)^2-s^2} \!\ \!\ \!\ 
\beta^{\dagger}_{+,{\bm b}+\frac{{\bm e}_{\mu}}{2}} 
\beta_{+,{\bm b}+\frac{{\bm e}_{\mu}}{2}} \nn \\
&\hspace{1cm} +  \sqrt{(E+t)^2-s^2} \!\ \!\ \!\ 
\beta^{\dagger}_{-,{\bm b}+\frac{{\bm e}_{\mu}}{2}} 
\beta_{-,{\bm b}+\frac{{\bm e}_{\mu}}{2}} \Big\}. \label{h0-app}
\end{align}

An evaluation based on 
eqs.~(\ref{rotation},\ref{t-ha},\ref{w1},\ref{eq1},\ref{full}) 
suggests that 
$t<0$ near (but above) 
the saturation field while $t>0$ in the high-field limit. The 
sign change is because the two-component 
vector $(u,v)$ is equally weighted in the particle-hole 
space near the saturation field (eq.~(\ref{eq1})), while it is 
fully polarized in the particle space in the high-field limit 
(eq.~(\ref{full})). In the present model, $t$ changes 
the sign around $H=1.05 H_c$, where the in-phase 
orbital level goes below the out-of-phase one 
in frequency.  Thus, in most of the fully polarized regime, 
we regard that the in-phase orbital 
at the $x$-link and that at the $y$-link  
comprises the lowest two.
    
In terms of the in-phase orbitals   
on the $x$-link and $y$-link, we next 
include ${\bm H}_1$ as inter-cluster transfer (hopping) 
integrals. To this end, we first describe ${\bm H}_1$, using 
the eigen basis of ${\bm H}^{\prime}_0$, 
$|u_{\mu,n,{\bm b}}\rangle$ ($n=0,1,2,3$ and $\mu=\pm$). 
The most dominant 
inter-cluster transfer integral is mainly from  
exchange processes between neighboring quadrants within 
the same ring. In terms of  
$\gamma^{\dagger}_{n,{\bm b}}$ and $\gamma_{n,{\bm b}}$, 
they are given by     
\begin{widetext}
\begin{align}
{\bm H}^{NN}_1 =& \sum_{{\bm b}} \bigg\{ 
\left(\begin{array}{cccc} 
\gamma^{\dagger}_{1,{\bm b}} & \gamma^{\dagger}_{0,{\bm b}} 
& \gamma_{1,{\bm b}} & \gamma_{0,{\bm b}} 
\end{array}\right)\left(\begin{array}{cccc} 
0 & \overline{A}_1 & 0 & \overline{B}_1 \\
\overline{A}^{*}_1 & 0 & \overline{B}_1 & 0 \\
0 & \overline{B}^{*}_1 & 0 & \overline{A}^{*}_1 \\
\overline{B}^{*}_1 & 0 & \overline{A}_1 & 0 \\
\end{array}\right) \left(\begin{array}{c}
\gamma_{1,{\bm b}} \\
\gamma_{0,{\bm b}} \\
\gamma^{\dagger}_{1,{\bm b}} \\
\gamma^{\dagger}_{0,{\bm b}} \\
\end{array}\right) \nn \\
& \hspace{1.0cm} 
+ \left(\begin{array}{cccc} 
\gamma^{\dagger}_{3,{\bm b}+{\bm e}_y} & \gamma^{\dagger}_{0,{\bm b}} 
& \gamma_{3,{\bm b}+{\bm e}_y} & \gamma_{0,{\bm b}} 
\end{array}\right)\left(\begin{array}{cccc} 
0 & \overline{A}_2 & 0 & \overline{B}_2 \\
\overline{A}^{*}_2 & 0 & \overline{B}_2 & 0 \\
0 & \overline{B}^{*}_2 & 0 & \overline{A}^{*}_2 \\
\overline{B}^{*}_2 & 0 & \overline{A}_2 & 0 \\
\end{array}\right) \left(\begin{array}{c}
\gamma_{3,{\bm b}+{\bm e}_y} \\
\gamma_{0,{\bm b}} \\
\gamma^{\dagger}_{3,{\bm b}+{\bm e}_y} \\
\gamma^{\dagger}_{0,{\bm b}} \\
\end{array}\right) \nn \\ 
&  \hspace{1.5cm} +  \left(\begin{array}{cccc} 
\gamma^{\dagger}_{2,{\bm b}+{\bm e}_x} & \gamma^{\dagger}_{1,{\bm b}} 
& \gamma_{2,{\bm b}+{\bm e}_x} & \gamma_{1,{\bm b}} 
\end{array}\right)\left(\begin{array}{cccc} 
0 & \overline{A}^{*}_2 & 0 & \overline{B}^{*}_2 \\
\overline{A}_2 & 0 & \overline{B}^{*}_2 & 0 \\
0 & \overline{B}_2 & 0 & \overline{A}_2 \\
\overline{B}_2 & 0 & \overline{A}^{*}_2 & 0 \\
\end{array}\right) \left(\begin{array}{c}
\gamma_{2,{\bm b}+{\bm e}_x} \\
\gamma_{1,{\bm b}} \\
\gamma^{\dagger}_{2,{\bm b}+{\bm e}_x} \\
\gamma^{\dagger}_{1,{\bm b}} \\
\end{array}\right) \nn \\ 
& \hspace{2.0cm} + \left(\begin{array}{cccc} 
\gamma^{\dagger}_{3,{\bm b}+{\bm e}_y} & 
\gamma^{\dagger}_{2,{\bm b}+{\bm e}_x} 
& \gamma_{3,{\bm b}+{\bm e}_y} & \gamma_{2,{\bm b}+{\bm e}_x} 
\end{array}\right) \left(\begin{array}{cccc} 
0 & \overline{A}_3 & 0 & \overline{B}_3 \\
\overline{A}^{*}_3 & 0 & \overline{B}_3 & 0 \\
0 & \overline{B}^{*}_3 & 0 & \overline{A}^{*}_3 \\
\overline{B}^{*}_3 & 0 & \overline{A}_3 & 0 \\
\end{array}\right) \left(\begin{array}{c}
\gamma_{3,{\bm b}+{\bm e}_y} \\
\gamma_{2,{\bm b}+{\bm e}_x} \\
\gamma^{\dagger}_{3,{\bm b}+{\bm e}_y} \\
\gamma^{\dagger}_{2,{\bm b}+{\bm e}_x} \\
\end{array}\right) \nn \\ 
& \hspace{2.0cm} 
+ \Big( \big\{{\bm e}_x,{\bm e}_y\big\},
\gamma^{(\dagger)}_{n,\cdots} 
\rightarrow  \big\{{\bm e}_{y}, -{\bm e}_x\big\}, 
\gamma^{(\dagger)}_{n+1,\cdots}
\Big) + \nn \\
&\hspace{1.0cm} 
+ \Big( \big\{{\bm e}_x,{\bm e}_y\big\},
\gamma^{(\dagger)}_{n,\cdots} 
\rightarrow  \big\{-{\bm e}_{x}, -{\bm e}_y\big\}, 
\gamma^{(\dagger)}_{n+2,\cdots} 
\Big)
+ \Big( \big\{{\bm e}_x,{\bm e}_y\big\},
\gamma^{(\dagger)}_{n,\cdots} 
\rightarrow  \big\{-{\bm e}_{y}, {\bm e}_x\big\}, 
\gamma^{(\dagger)}_{n+3,\cdots} \Big)
\bigg\}, \label{nn1}
\end{align}
\end{widetext}
with 
\begin{align} 
\overline{A}_1 & \equiv \langle u_{+,1,{\bm b}} |  
{\bm H}_{1} | u_{+,0,{\bm b}} \rangle, \label{a1-ha} \\
 \overline{B}_1 &\equiv \langle u_{+,1,{\bm b}} | 
{\bm H}_{1}| u_{-,0,{\bm b}} \rangle, \label{b1-ha}
\end{align}
\begin{align}
\overline{A}_2 & \equiv \langle 
u_{+,3,{\bm b}+{\bm e}_y} |
{\bm H}_{1} | u_{+,0,{\bm b}} \rangle, \label{a2-ha} \\
 \overline{B}_2 &\equiv \langle u_{+,3,{\bm b}+{\bm e}_y} 
| {\bm H}_{1} | u_{-,0,{\bm b}} \rangle, \label{b2-ha} 
\end{align}
\begin{align}
\overline{A}_3 &\equiv \langle 
u_{+,3,{\bm b}+{\bm e}_y} | 
{\bm H}_{1} | u_{+,2,{\bm b}+{\bm e}_x} \rangle, \label{a3-ha} \\ 
 \overline{B}_3 &\equiv \langle u_{+,3,{\bm b}+{\bm e}_y} | 
{\bm H}_{1} | u_{-,2,{\bm b}+{\bm e}_x} \rangle. \label{b3-ha} 
\end{align} 
The 2nd line and 3rd line in the r.h.s. of eq.~(\ref{nn1}) 
are related to each other by a combined symmetry between 
the time-reversal and a in-plane mirror which interchanges 
$x$-axis and $y$-axis. 

The next dominant inter-cluster transfer integrals 
are between the next-nearest-neighbor (NNN) clusters 
and they hvae two kinds; one is 
$(\sigma,\sigma)$-coupling type, which is 
between two in-phase orbitals 
on $x$-links connected by ${\bm e}_x$ or those 
on $y$-links connected by ${\bm e}_y$ 
(Fig.~\ref{fig:2-orbital-TB}). 
The other is $(\pi,\pi)$-coupling 
type, which is between two in-phase orbitals 
on $x$-links connected by ${\bm e}_y$ or 
those on the $y$-links connected by ${\bm e}_x$ 
(Fig.~\ref{fig:2-orbital-TB}). They are given by        
\begin{widetext}
\begin{align}
{\bm H}^{NNN,\sigma\sigma}_1 = & \sum_{{\bm b}} \bigg\{ 
\left(\begin{array}{cccc} 
\gamma^{\dagger}_{2,{\bm b}} & \gamma^{\dagger}_{0,{\bm b}} 
& \gamma_{2,{\bm b}} & \gamma_{0,{\bm b}} 
\end{array}\right)\left(\begin{array}{cccc} 
0 & \overline{C}_1 & 0 & \overline{D}_1 \\
\overline{C}^{*}_1 & 0 & \overline{D}_1 & 0 \\
0 & \overline{D}^{*}_1 & 0 & \overline{C}^{*}_1 \\
\overline{D}^{*}_1 & 0 & \overline{C}_1 & 0 \\
\end{array}\right) \left(\begin{array}{c}
\gamma_{2,{\bm b}} \\
\gamma_{0,{\bm b}} \\
\gamma^{\dagger}_{2,{\bm b}} \\
\gamma^{\dagger}_{0,{\bm b}} \\
\end{array}\right) \nn \\
&  \hspace{1cm} + \left(\begin{array}{cccc} 
\gamma^{\dagger}_{2,{\bm b}} & \gamma^{\dagger}_{2,{\bm b}+{\bm e}_x} 
& \gamma_{2,{\bm b}} & \gamma_{2,{\bm b}+{\bm e}_x} 
\end{array}\right)\left(\begin{array}{cccc} 
0 & \overline{C}_2 & 0 & \overline{D}_2 \\
\overline{C}^{*}_2 & 0 & \overline{D}_2 & 0 \\
0 & \overline{D}^{*}_2 & 0 & \overline{C}^{*}_2 \\
\overline{D}^{*}_2 & 0 & \overline{C}_2 & 0 \\
\end{array}\right) \left(\begin{array}{c}
\gamma_{2,{\bm b}} \\
\gamma_{2,{\bm b}+{\bm e}_x} \\
\gamma^{\dagger}_{2,{\bm b}} \\
\gamma^{\dagger}_{2,{\bm b}+{\bm e}_x} \\
\end{array}\right) \nn \\
&  \hspace{1.5cm} + \left(\begin{array}{cccc} 
\gamma^{\dagger}_{0,{\bm b}} & \gamma^{\dagger}_{0,{\bm b}-{\bm e}_x} 
& \gamma_{0,{\bm b}} & \gamma_{0,{\bm b}-{\bm e}_x} 
\end{array}\right)\left(\begin{array}{cccc} 
0 & \overline{C}_2 & 0 & \overline{D}_2 \\
\overline{C}^{*}_2 & 0 & \overline{D}_2 & 0 \\
0 & \overline{D}^{*}_2 & 0 & \overline{C}^{*}_2 \\
\overline{D}^{*}_2 & 0 & \overline{C}_2 & 0 \\
\end{array}\right) \left(\begin{array}{c}
\gamma_{0,{\bm b}} \\
\gamma_{0,{\bm b}-{\bm e}_x} \\
\gamma^{\dagger}_{0,{\bm b}} \\
\gamma^{\dagger}_{0,{\bm b}-{\bm e}_x} \\
\end{array}\right) \nn \\ 
&  \hspace{2.0cm} + \left(\begin{array}{cccc} 
\gamma^{\dagger}_{2,{\bm b}+{\bm e}_x} 
& \gamma^{\dagger}_{0,{\bm b}-{\bm e}_x} 
& \gamma_{2,{\bm b}+{\bm e}_x} & \gamma_{0,{\bm b}-{\bm e}_x} 
\end{array}\right)\left(\begin{array}{cccc} 
0 & \overline{C}_3 & 0 & \overline{D}_3 \\
\overline{C}^{*}_3 & 0 & \overline{D}_3 & 0 \\
0 & \overline{D}^{*}_3 & 0 & \overline{C}^{*}_3 \\
\overline{D}^{*}_3 & 0 & \overline{C}_3 & 0 \\
\end{array}\right) \left(\begin{array}{c}
\gamma_{2,{\bm b}+{\bm e}_x} \\
\gamma_{0,{\bm b}-{\bm e}_x} \\
\gamma^{\dagger}_{2,{\bm b}+{\bm e}_x} \\
\gamma^{\dagger}_{0,{\bm b}-{\bm e}_x} \\
\end{array}\right) \nn \\ 
& \hspace{2.5cm} + \Big({\bm e}_x, 
\gamma^{(\dagger)}_{0,\cdots}, 
\gamma^{(\dagger)}_{2,\cdots} 
\rightarrow {\bm e}_y, 
\gamma^{(\dagger)}_{1,\cdots}, 
\gamma^{(\dagger)}_{3,\cdots} \Big) 
\bigg\}, \label{nnns}
\end{align}
and
\begin{align}
{\bm H}^{NNN,\pi\pi}_1 = & \sum_{{\bm b}} \bigg\{ 
\left(\begin{array}{cccc} 
\gamma^{\dagger}_{0,{\bm b}+{\bm e}_y} 
& \gamma^{\dagger}_{0,{\bm b}} 
& \gamma_{0,{\bm b}+{\bm e}_y} & \gamma_{0,{\bm b}} 
\end{array}\right)\left(\begin{array}{cccc} 
0 & \overline{E}_1 & 0 & \overline{F}_1 \\
\overline{E}^{*}_1 & 0 & \overline{F}_1 & 0 \\
0 & \overline{F}^{*}_1 & 0 & \overline{E}^{*}_1 \\
\overline{F}^{*}_1 & 0 & \overline{E}_1 & 0 \\
\end{array}\right) \left(\begin{array}{c}
\gamma_{0,{\bm b}+{\bm e}_y} \\
\gamma_{0,{\bm b}} \\
\gamma^{\dagger}_{0,{\bm b}+{\bm e}_y} \\
\gamma^{\dagger}_{0,{\bm b}} \\
\end{array}\right) \nn \\
&  \hspace{1cm} + \left(\begin{array}{cccc} 
\gamma^{\dagger}_{2,{\bm b}+{\bm e}_x+{\bm e}_y} 
& \gamma^{\dagger}_{2,{\bm b}+{\bm e}_x} 
& \gamma_{2,{\bm b}+{\bm e}_x+{\bm e}_y} 
& \gamma_{2,{\bm b}+{\bm e}_x} 
\end{array}\right)\left(\begin{array}{cccc} 
0 & \overline{E}^{*}_1 & 0 & \overline{F}_1 \\
\overline{E}_1 & 0 & \overline{F}_1 & 0 \\
0 & \overline{F}^{*}_1 & 0 & \overline{E}_1 \\
\overline{F}^{*}_1 & 0 & \overline{E}^{*}_1 & 0 \\
\end{array}\right) \left(\begin{array}{c}
\gamma_{2,{\bm b}+{\bm e}_x+{\bm e}_y} \\
\gamma_{2,{\bm b}+{\bm e}_x} \\
\gamma^{\dagger}_{2,{\bm b}+{\bm e}_x+{\bm e}_y} \\
\gamma^{\dagger}_{2,{\bm b}+{\bm e}_x} \\
\end{array}\right) \nn \\
& \hspace{1.5cm} + \left(\begin{array}{cccc} 
\gamma^{\dagger}_{2,{\bm b}+{\bm e}_x+{\bm e}_y} 
& \gamma^{\dagger}_{0,{\bm b}} 
& \gamma_{2,{\bm b}+{\bm e}_x+{\bm e}_y} 
& \gamma_{0,{\bm b}} 
\end{array}\right)\left(\begin{array}{cccc} 
0 & \overline{E}_2 & 0 & \overline{F}_2 \\
\overline{E}_2 & 0 & \overline{F}_2 & 0 \\
0 & \overline{F}^{*}_2 & 0 & \overline{E}_2 \\
\overline{F}^{*}_2 & 0 & \overline{E}_2 & 0 \\
\end{array}\right) \left(\begin{array}{c}
\gamma_{2,{\bm b}+{\bm e}_x+{\bm e}_y} \\
\gamma_{0,{\bm b}} \\
\gamma^{\dagger}_{2,{\bm b}+{\bm e}_x+{\bm e}_y} \\
\gamma^{\dagger}_{0,{\bm b}} \\
\end{array}\right) \nn \\
& \hspace{2.0cm} + \left(\begin{array}{cccc} 
\gamma^{\dagger}_{0,{\bm b}+{\bm e}_y} 
& \gamma^{\dagger}_{2,{\bm b}+{\bm e}_x} 
& \gamma_{0,{\bm b}+{\bm e}_y} 
& \gamma_{2,{\bm b}+{\bm e}_x} 
\end{array}\right)\left(\begin{array}{cccc} 
0 & \overline{E}_2 & 0 & \overline{F}^{*}_2 \\
\overline{E}_2 & 0 & \overline{F}^{*}_2 & 0 \\
0 & \overline{F}_2 & 0 & \overline{E}_2 \\
\overline{F}_2 & 0 & \overline{E}_2 & 0 \\
\end{array}\right) \left(\begin{array}{c}
\gamma_{0,{\bm b}+{\bm e}_y} \\
\gamma_{2,{\bm b}+{\bm e}_x} \\
\gamma^{\dagger}_{0,{\bm b}+{\bm e}_y} \\
\gamma^{\dagger}_{2,{\bm b}+{\bm e}_x} \\
\end{array}\right) \nn \\
& \hspace{2.5cm} + \Big({\bm e}_x, {\bm e}_y,
\gamma_{0,\cdots},\gamma^{\dagger}_{0,\cdots},  
\gamma_{2,\cdots},\gamma^{\dagger}_{2,\cdots},  
\rightarrow {\bm e}_y, {\bm e}_x,
\gamma^{\dagger}_{1,\cdots},\gamma_{1,\cdots}, 
\gamma^{\dagger}_{3,\cdots},\gamma_{3,\cdots} \Big)
 \bigg\},  \label{nnnp}
\end{align}
\end{widetext}
with 
\begin{align}
\overline{C}_1 & \equiv \langle u_{+,2,{\bm b}} |
{\bm H}_{1} | u_{+,0,{\bm b}} \rangle, \label{c1-ha} \\
\overline{D}_1 &= \langle u_{+,2,{\bm b}} |  
{\bm H}_{1} | u_{-,0,{\bm b}} \rangle, \label{d1-ha}
\end{align}
\begin{align}
\overline{C}_2 & 
\equiv \langle u_{+,2,{\bm b}} | 
{\bm H}_{1} | u_{+,2,{\bm b}+{\bm e}_x} \rangle, \label{c2-ha} \\
\overline{D}_2 &= \langle u_{+,2,{\bm b}} |  
{\bm H}_{1} | u_{-,2,{\bm b}+{\bm e}_x} \rangle, \label{d2-ha} 
\end{align}
\begin{align}
\overline{C}_3 & \equiv \langle u_{+,2,{\bm b}+{\bm e}_x} | 
{\bm H}_{1}  | u_{+,0,{\bm b}-{\bm e}_x} \rangle, \label{c3-ha} \\
\overline{D}_3 &= \langle u_{+,2,{\bm b}+{\bm e}_x} |  
{\bm H}_{1} | u_{-,0,{\bm b}-{\bm e}_x} \rangle, \label{d3-ha} 
\end{align}
\begin{align}
\overline{E}_1 & \equiv \langle 
u_{+,0,{\bm b}+{\bm e}_y} | 
{\bm H}_{1} | u_{+,0,{\bm b}} \rangle, \label{e1-ha} \\
\overline{F}_1 &= \langle u_{+,0,{\bm b}+{\bm e}_y} | 
 {\bm H}_{1} | u_{-,0,{\bm b}} \rangle, \label{f1-ha}. 
\end{align}
\begin{align}
\overline{E}_2 & \equiv \langle 
u_{+,2,{\bm b}+{\bm e}_x+{\bm e}_y} |
{\bm H}_{1} | u_{+,0,{\bm b}} \rangle, \label{e2-ha} \\
\overline{F}_2 &= \langle 
u_{+,2,{\bm b}+{\bm e}_x+{\bm e}_y} |
{\bm H}_{1} | u_{-,0,{\bm b}} \rangle. \label{f2-ha} 
\end{align}
Evaluations based on 
eqs.~(\ref{rotation},\ref{p-h},\ref{w1},\ref{full},\ref{a1-ha}-\ref{f2-ha})  
suggests that $\overline{A}_1=ia_1$,   
$\overline{A}_2= -ia_2$, $\overline{A}_3=ia_3$, $\overline{B}_1
=b_1$, $\overline{B}_2=-b_2$, $\overline{B}_3=b_3$, 
$\overline{C}_1=c_1$, $\overline{C}_2=-c_2$, 
$\overline{C}_3=c_3$, $\overline{D}_1=d_1$, $\overline{D}_2=-d_2$, 
$\overline{D}_3=d_3$, $\overline{E}_1=-e_1$, $\overline{E}_2=e_2$, 
$\overline{F}_1= f_1$, $\overline{F}_2 = - f_2$ with 
real and positive $a_1$, $a_2$, $a_3$ 
($a_1\gtrsim a_2 \gtrsim a_3>0$), $b_1$, $b_2$, $b_3$  
($b_1\gtrsim b_2 \gtrsim b_3>0$), $c_1$, $c_2$, $c_3$ 
($c_1\gtrsim c_2 \gtrsim c_3>0$), $d_1$, $d_2$, $d_3$ 
($d_1\gtrsim d_2 \gtrsim d_3>0$), $e_1$, $e_2$, $f_1$, $f_2$, 
($e_1 \gtrsim e_2$) and ($f_1 \gtrsim f_2$). 

Using eqs.~(\ref{bog}), we rewrite 
Eqs.~(\ref{nn1},\ref{nnns},\ref{nnnp}) in the basis 
of the in-phase ($\beta_{+,{\bm b}+\frac{{\bm e}_x}{2}}$, 
$\beta_{+,{\bm b}+\frac{{\bm e}_y}{2}}$) and out-of-phase  
($\beta_{-,{\bm b}+\frac{{\bm e}_x}{2}}$, 
$\beta_{-,{\bm b}+\frac{{\bm e}_y}{2}}$) orbital wave 
functions. In most of the fully polarized regime, the 
in-phase orbital level goes below the out-of-phase orbital level. %, 
%$\sqrt{(E+t)^2-s^2}>\sqrt{(E-t)^2-s^2}$. 
Focusing on the lowest two volume-mode bands, we   
thus ignore those transfer integrals which are involved with 
out-of-phase orbitals. This leads to,   
\begin{align}
\overline{\bm H}=\sum_{\bm b}\big\{ 
\overline{\bm H}_0 
+ \overline{\bm H}^{NN}_1 + 
\overline{\bm H}^{NNN}_1\big\}, \label{app-total}
\end{align} 
\begin{align}
\overline{\bm H}_0 &= \Delta \!\  
\beta^{\dagger}_{+,{\bm b}+\frac{{\bm e}_x}{2}} 
\beta_{+,{\bm b}+\frac{{\bm e}_x}{2}} 
+ \Delta \!\ \beta^{\dagger}_{+,{\bm b}+\frac{{\bm e}_y}{2}} 
\beta_{+,{\bm b}+\frac{{\bm e}_y}{2}}  \label{app-h0}
\end{align}
\begin{align}
\overline{\bm H}^{NN}_{1} &=  (i a + b) \!\ 
\beta^{\dagger}_{+,{\bm b}+\frac{{\bm e}_y}{2}} 
\beta_{+,{\bm b}+\frac{{\bm e}_x}{2}} 
\nn \\
&\ \ \ - (i a+b) \!\ 
\beta^{\dagger}_{+,{\bm b}+{\bm e}_y+\frac{{\bm e}_x}{2}} 
\beta_{+,{\bm b}+\frac{{\bm e}_y}{2}} \nn \\
& \ \ \ \ + (i a + b) \!\ 
\beta^{\dagger}_{+,{\bm b}+{\bm e}_x + \frac{{\bm e}_y}{2}} 
\beta_{+,{\bm b}+{\bm e}_y+\frac{{\bm e}_x}{2}} \nn \\
& \hspace{0.3cm} - (i a+ b) \!\ 
\beta^{\dagger}_{+,{\bm b}+\frac{{\bm e}_x}{2}} 
\beta_{+,{\bm b}+{\bm e}_x+\frac{{\bm e}_y}{2}} + {\rm h.c.} 
\label{app-h1}
\end{align}  
\begin{align}
\overline{\bm H}^{NNN}_1 &= c \!\ 
\beta^{\dagger}_{+,{\bm b}+\frac{{\bm e}_x}{2}} 
\beta_{+,{\bm b} - \frac{{\bm e}_x}{2}} 
+ c \!\ \beta^{\dagger}_{+,{\bm b}+\frac{{\bm e}_y}{2}} 
\beta_{+,{\bm b} - \frac{{\bm e}_y}{2}} + \nn \\
& \hspace{-1.2cm}  c^{\prime} \!\ 
\beta^{\dagger}_{+,{\bm b}+{\bm e}_y +\frac{{\bm e}_x}{2}} 
\beta_{+,{\bm b} + \frac{{\bm e}_x}{2}} 
+ c^{\prime} \!\ 
\beta^{\dagger}_{+,{\bm b}+{\bm e}_x 
+\frac{{\bm e}_y}{2}} 
\beta_{+,{\bm b} + \frac{{\bm e}_y}{2}} + {\rm h.c.}, \label{app-h2} 
\end{align}
with 
\begin{align}
\Delta & = \sqrt{(E-t)^2-s^2} \\
a &= a_1+2a_2+a_3, \label{a-ha} \\
b  &= 
\frac{(b_1+2b_2+b_3)\cdot s}{\sqrt{(E-t)^2-s^2}}, \label{b-ha} \\
c & = - \frac{(d_1+2d_2+d_3)\cdot s 
+ (c_1+2c_2+c_3)\cdot (E-t) }{\sqrt{(E-t)^2-s^2}}, \label{c-ha} \\ 
c^{\prime} & = \frac{2(f_1+f_2)\cdot s 
- 2(e_1+e_2)\cdot (E-t)}{\sqrt{(E-t)^2-s^2}}. \label{cd-ha} 
\end{align}
Since the particle space and the hole space 
is separated by a large frequency spacing, 
$2\Delta$, we have also omitted 
hopping terms in particle-particle channel, 
such as $\beta^{\dagger} \beta^{\dagger}$ 
and $\beta \beta$.  
$a \!\ (>0)$ and $b \!\ (>0)$ quantify an imaginary 
part and real part of the nearest-neighbor 
inter-cluster transfer integral, while $c \!\ (<0)$ 
and $c^{\prime} \!\ (<0)$ stand for the $(\sigma,\sigma)$-coupling 
and the $(\pi,\pi)$-coupling next-nearest-neighbor 
transfer integrals respectively. 
An amplitude of transfer integral is  
inversely proportional to the cubic in distance 
(eq.~(\ref{rotation})), so that the 
$(\sigma,\sigma)$-coupling type is expected to be larger 
than the $(\pi,\pi)$-coupling type, 
$|c|>|c^{\prime}|$ (or $c_1+2c_2+c_3>2e_1+2e_2$). 
Note also that $b\rightarrow 0$ in the limit of $H\rightarrow +\infty$, 
where $t/E,s/E\rightarrow 0$. By replacing 
$\beta_{+,{\bm b}+\frac{{\bm e}_x}{2}}$ and 
$\beta_{+,{\bm b}+\frac{{\bm e}_y}{2}}$ by 
$\beta_{{\bm b}+\frac{{\bm e}_x}{2}}$ and 
$\beta_{{\bm b}+\frac{{\bm e}_y}{2}}$ respectively, 
we have eqs.~(\ref{h0},\ref{h1}).    


\begin{thebibliography}{99}
\bibitem{YIG} A. A. Serga, 
A. V. Chumak and B. Hillebrands, J. Phys. D: Appl. Phys. \textbf{43},  
264002 (2010). 
\bibitem{KDG} V. V. Kruglyak, S. O. Demokritov, and D. Grundler, 
J. Phys. D, \textbf{43}, 264001 (2010). 
\bibitem{DE1} R. W. Damon and J. R. Eshbach, J. Phys. Chem. Solids, 
\textbf{19}, 308 (1961).
\bibitem{Kostylev} M. P. Kostylev, A. A. Serga, 
T. Schneider, B. Leven, and B. Hillebrands, 
Appl. Phys. Lett. \textbf{87}, 153501 (2005). 
\bibitem{Lee} K. S. Lee and S. K. Kim, J. Appl. Phys. \textbf{104},  
053909 (2008). 
\bibitem{Schneider} T. Schneider, A. A. Serga, B. Leven, 
B. Hillebrands, R. L. Stamps, and M. P.  Kostylev, 
Appl. Phys. Lett. \textbf{92}, 022505 (2008). 
\bibitem{Sato} N. Sato, K. Sekiguchi, Y. Nozaki, 
Appl. Phys. Express. \textbf{6}, 063001 (2013).    
\bibitem{SO1} R. Shindou, R. Matsumoto, S. Murakami, 
and J-i Ohe, Phys. Rev. B, \textbf{87}, 174427 (2013). 
\bibitem{SO2} R. Shindou, J-i Ohe, R. Matsumoto, 
S. Murakami, and E. Saitoh, Phys. Rev. B, \textbf{87}, 174402 (2013).
\bibitem{TKKN} D. J. Thouless, M. Kohmoto, M. P. Nightingale, 
and M. den Nijs, Phys. Rev. Lett. \textbf{49}, 405 (1982).
\bibitem{Hal} B. I. Halperin, Phys. Rev. B \textbf{25}, 2185 (1982).
\bibitem{Hat} Y. Hatsugai, Phys. Rev. Lett. \textbf{71}, 3697 (1993).
\bibitem{DV2} R. W. Damon and H. Van De Varrt, J. Appl. Phys. \textbf{36}, 
3453 (1965). 
\bibitem{Kalinikos} B. A. Kalinikos, and A. N. Slavin, J. Phys. C: Solid State Phys 
\textbf{19}, 7013 (1986). 
\bibitem{AM} R. Arias, and D. L. Mills, 
Phys. Rev. B, \textbf{63}, 134439 (2001).
\bibitem{Volovik} G. E. Volovik, Sov. 
Phys. JETP, \textbf{67}, 1804 (1988). 
\bibitem{Yakovenko} V. M. Yakovenko, 
Phys. Rev. Letters, \textbf{65}, 251 (1990). 
\bibitem{QWZ} X. L. Qi, Y. S. Wu, and 
S. C. Zhang, Phys. Rev. B \textbf{74}, 085308 (2006).  
\bibitem{BTZ} B. A. Bernevig, T. L. Hughes, and S. C. Zhang, 
Science \textbf{314}, 1757 (2006). 
\bibitem{FK} L. Fu and C. L. Kane, Phys. Rev. B \textbf{76}, 045302 (2007).    
\bibitem{Adeyeye} A. O. Adeyeye and N. Singh, J. Phys. D: Appl. Phys. 
\textbf{41}, 153001 (2008).
\bibitem{Gulyaev} Y. V. Gulyaev, JETP Lett. \textbf{77}, 567 (2003). 
\bibitem{Hubert} A. Hubert, and R. Schafer, 
{\it Magnetic Domains} (Springer, Berlin, Germany, 2000).
\bibitem{Qiu} Z. Q. Qiu, J. Pearson, S. D. Bader, Phys. Rev. Letters, 
\textbf{70}, 1006 (1993). 
\bibitem{Cowburn} R. P. Cowburn, D. K. Koltsov, 
A. O. Adeyeye, M. E.  Welland, and D. M. Tricker, 
Phys. Rev. Letters, \textbf{83}, 1042 (1999). 
\bibitem{Shinjo} T. Shinjo, T. Okuno, R. Hassdorf, and K. Shigeto, 
and T. Ono, Science, \textbf{289}, 930 (2000).  
\bibitem{Avron} J. E. Avron, R. Seiler and B. Simon, Phys. Rev. Letters, 
\textbf{51}, 51 (1983).  
\end{thebibliography}
\end{document}